\documentclass[iop]{emulateapj}
\usepackage{amsmath}
\renewcommand{\sun}{0}

\usepackage{color}

\shorttitle{Mergers by Precession-induced Resonances}
\shortauthors{Bhaskar et al.}

\begin{document}

\title{Enhanced Blackhole mergers in AGN discs due to Precession induced resonances }

\author{Hareesh Gautham Bhaskar \altaffilmark{1}, Gongjie Li \altaffilmark{1}, Douglas Lin \altaffilmark{2}}
\affil{$^1$ Center for Relativistic Astrophysics, School of Physics, Georgia Institute of Technology, Atlanta, GA 30332, USA}
\affil{$^2$ University of California , Santa Cruz}


\begin{abstract}
Recent studies have shown that AGN discs can host sources of gravitational waves. Compact binaries can form and merge in AGN discs through their interactions with the gas and other compact objects in the disc. It is also possible for the binaries to shorten the merging timescale due to eccentricity excitation caused by perturbations from the supermassive blackhole (SMBH). In this paper we focus on effects due to precession-induced (eviction-like) resonances, where nodal and apsidal precession rates of the binary is commensurable with the mean motion of the binary around the SMBH. We focus on intermediate mass black hole (IMBH)-stellar mass black hole (SBH) binaries, and consider binary orbit inclined from the circum-IMBH disk which leads to the orbital $J_2$ precession. We show that if a binary is captured in these resonances and is migrating towards the companion, it can undergo large eccentricity and inclination variations. We derive analytical expressions for the location of fixed points, libration timescale and width for these resonances, and identified two resonances in the near coplanar regime (the evection and eviction resonances) as well as two resonances in the near polar regime that can lead to mergers. We also derive analytical expressions for the maximum eccentricity that a migrating binary can achieve for given initial conditions. Specifically, the maximum eccentricity can reach 0.9 when captured in these resonances before orbital decay due to gravitational wave emission dominates, and the capture is only possible for slow migration ($\sim 10$ Myr) 2-3 order of magnitude longer than the resonance libration timescale. We also show that capture into multiple resonances is possible, and can further excite eccentricities.
\end{abstract}


\section{Introduction}

Multiple mechanisms have been proposed to account for the recent detections of gravitational waves. It has been suggested that the sources for these events could be merging compact objects in isolated binaries \citep{dominik_double_2012,kinugawa_possible_2014,belczynski_first_2016,belczynski_origin_2018,giacobbo_merging_2018,spera_merging_2019,bavera_origin_2020}, isolated triple systems \citep{Li14_flip, hoang_black_2018,antonini_black_hole_2014,silsbee_lidov-kozai_2017,antonini_binary_2017} and star clusters \citep{banerjee_stellar-mass_2017,oleary_binary_2006,samsing_formation_2014,rodriguez_binary_2016,askar_mocca-survey_2017,zevin_eccentric_2019,di_carlo_merging_2019}. Additionally, AGN discs which are dense with embedded stars and compact objects are also expected to be sources of gravitational waves. A wide variety of processes can aid in the merger of compact objects in AGN discs including flybys and gas dynamical friction \citep{bartos_rapid_2017,tagawa_formation_2020,leigh_rate_2018,stone_assisted_2017,tagawa_eccentric_2021}.

More recently, it has been shown that migrating binary black holes in an AGN disc can be captured in evection resonances causing the eccentricity of the binary to be excited allowing the binaries to merge on a shorter timescale \citep{munoz_eccentric_2022,bhaskar_black_2022}. In this work, we explore other resonances which can also cause similar eccentricity excitation. In our previous paper \citep{bhaskar_black_2022}, we assumed a coplanar setup which allows resonances involving only the longitude of pericenter of the binary. In this work we relax that assumption allowing the binaries to have an arbitrary orientation with respect to the AGN disc, and for simplicity we now focus on IMBH-SBH binaries.

More specifically, in this paper we focus on resonances which occur when a linear combination of apsidal and nodal precession rates of the binary is commensurable with the mean motion of the companion. The most well studied of these resonances is the evection resonance which occurs when the precession rate of the longitude of pericenter of the binary equals the mean motion of the companion. Evection resonances have been shown to be important in many astrophysical systems. It was initially used to show that the Moon in it's early evolution could have been trapped in evection resonance induced by the Sun, allowing the transfer of angular momentum from Earth-Moon orbit to Earth-Sun orbit \citep{touma_resonances_1998,cuk_dynamical_2019, tian2017}. More recently, it has been shown that circumbinary planets with an external perturber \citep{xu_disruption_2016}, binary blackholes in AGN discs \citep{bhaskar_black_2022, munoz_eccentric_2022}, multiplanet systems \citep{touma_disruption_2015} and satellites around exoplanets \citep{spalding_resonant_2016,cuk_secular_2004} can be trapped in evection resonance. Other precession induced resonances have also been studied. For instance, \cite{touma_resonances_1998} show that the moon could also have been trapped in Eviction resonance in the past. Additionally, \cite{yoder_tidal_1982}, \cite{yokoyama_possible_2002} and \cite{vaillant_eviction-like_2022} explore the possibility of capture of moons of Mars and Neptune in other precession induced resonances .

Our main interest in studying these resonances is to explore the possibility of eccentricity excitation in binaries.  While mechanisms which allow eccentricity excitation are of interest to many astrophysical systems, in this paper we focus on blackhole binaries in AGN disc. The paper is organized as follows: In \S \ref{sec:antheory}, we develop the analytical theory for the system. We derive the Hamiltonian describing the dynamics of the system. In \S \ref{sec:pirestheory}, we focus on the dynamics of binaries trapped in precession induced resonances. We derive analytical expressions for resonance location, libration timescale and resonant width at low eccentricity. We then describe our single-averaged simulations in \S \ref{sec:secsim}. We apply our results to BBHs in AGN discs in \S \ref{sec:bbhagn}. We summarize our results and conclude in \S \ref{sec:conc}. Readers only interested in applications to AGN disks can skip \S \ref{sec:antheory} and \S \ref{sec:secsim}.

\section{Analytical theory} 

\label{sec:antheory}
\subsection{Single averaged Hamiltonian}

\begin{figure}
	\centering
	\includegraphics[height=0.8\linewidth,width=1.0\linewidth]{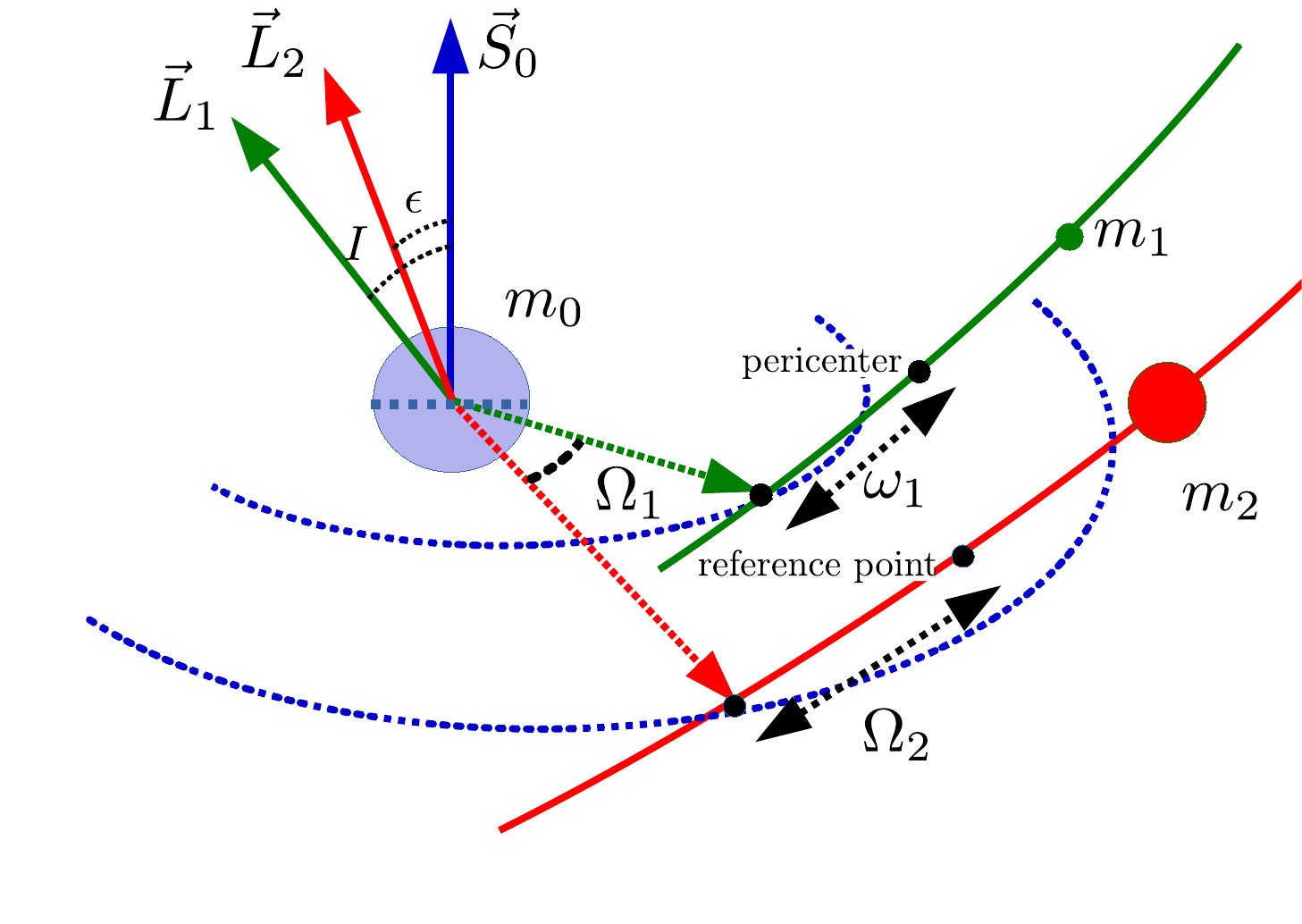}
	\caption{The schematic of the system. It shows a binary comprising of masses $m_0$ and $m_1$ being orbited by mass $m_2$. The inclination of the inner orbit ($I$) and the outer orbit ($\epsilon$) are measured with respect to the angular momentum of the disk denoted by $\vec{S}_0$. The longitude of ascending node of the inner and the outer binaries are denoted by $\Omega_1$ and $\Omega_2$ respectively and the argument of pericenter of the inner orbit is denoted by $\omega_1$. }
	\label{fig:schematic}
\end{figure}
The schematic of the system is shown in Figure \ref{fig:schematic}. It contains a binary comprising of objects with masses $m_0$ and $m_1$ being orbited by a third object with mass $m_2$ on a wider orbit. We define an inner orbit tracking the evolution of masses $m_0$ and $m_1$ around each other and an outer orbit tracking the evolution of the outer companion ($m_2$) around the center of mass of the inner binary. The orbital elements of the inner (outer) orbit are defined as follows: $a_1$ ($a_2$) is the semi-major axis, $e_1$ ($e_2$) is the eccentricity, $I$ ($\epsilon$) is the inclination, $\omega_1$ ($\omega_2$) is the argument of pericenter, $\Omega_1$($\Omega_2$) is the longitude of ascending node and $\lambda_1$($\lambda_2$) is the mean longitude. Subscript ``1'' denotes the orbit of $m_1$ around $m_0$ (inner binary), and ``2'' denotes the orbit of $m_2$ around the center of mass of $m_0$ and $m_1$ (outer binary). To simplify the analysis, we take the eccentricity of the outer orbit to be zero. In addition, without loss of generality we set the longitude of pericenter of the outer companion to zero. We follow the dynamics of the inner binary, by including following effects: precession due to disc around mass $m_0$ (modeled using a $J_2$ term) and perturbations from the outer companion. For simplicity from this point on, $e$, $\omega$ and $\Omega$ implicitly implies $e_1$, $\omega_1$ and $\Omega_1$ respectively. The Hamiltonian of the system can be written as
\begin{equation}
	H=H_{J_2}+H_{PN}+H_{m_2} \label{eq:hamcmp}
\end{equation}
where the first, second and third terms represent contributions from the quadrupole moment of the disc, post-Newtonian effects and perturbations from the companion respectively. We assume that the system is hierarchical ($a_1 \ll a_2$) and that the outer companion is much more massive than either components of the inner binary. It follows from these assumptions that the outer orbit is constant.  To simplify the analysis, the interaction potential of the companion ($H_{m_2}$) is expanded upto quadrupole order in the ratio of semi-major axis of the inner and the outer orbits ($a_1/a_2$), with hexadecapole and higher order terms assumed to be negligible \footnote{For systems with perturbers on circular orbits, the octupole order term vanishes. Under certain conditions, hexadecapole and higher order terms can be important \citep{will_orbital_2017}.}. In this work we mainly focus on dynamical timescales much longer than the orbital period of the inner binary. Hence, the Hamiltonian is averaged over the shortest timescale of the systems, which is the orbital period of the inner binary. The averaged Hamiltonian for the quadrupole potential and post-Newtonian terms is given by:
\begin{eqnarray}
H_{J_2} &=& -\frac{GJ_2R_d^2 m_0 }{8a_1^3 (1-e^2)^{3/2}}(1+3\cos{2I}), \nonumber \\
H_{PN} &=& -\frac{3G^2(m_0+m_1)^2}{a_1^2c^2}\frac{1}{\sqrt{1-e^2}}, \nonumber
\end{eqnarray}
where $G$ is the universal gravitational constant and $c$ is the speed of light  \citep{murray_solar_1999,touma_resonances_1998,eggleton_orbital_2001}. The single averaged Hamiltonian for the companion is more complicated and contains several terms which represent resonances in linear combinations of argument of pericenter, longitude of ascending node of the inner binary and mean longitude of the outer binary. { A simplified expression of the single-averaged quadrupole order Hamiltonian can be found in \cite{vaillant_eviction-like_2022} (See Eqn. 13). Also, the Hamiltonian is truncated such that the terms which are negligible in the context of this study have been dropped}.  The expression for the Hamiltonian can be written as:
\begin{equation}
    H_{m_2} = -\frac{Gm_2}{a_2} \left(\frac{a_1}{a_2}\right)^2 \sum_i S_{\phi_i}(I,\epsilon,e)\cos(\phi_i)
\end{equation}
where $\phi_i$ are linear combinations of arguments of pericenter and longitudes of ascending node of the inner and the outer orbits and mean longitude of the outer orbit. In addition, $S_{\phi_i}(I,\epsilon,e)$ are functions of inclination, eccentricity and obliquity whose form depends on the resonant angle $\phi_i$.  For instance, the terms corresponding to evection and eviction resonances are given by:
\begin{eqnarray}
H_{evection} &=& -f_s (1+\cos{I})^2(1+\cos{\epsilon})^2 \cos(2f_2-2\omega-2\Omega) \nonumber \\
H_{eviction} &=&  -16 f_s \sin{I}\cos^2{\frac{I}{2}}\cos^2{\frac{\epsilon}{2}}\sin{\epsilon} \cos(2f_2-2\omega-\Omega) \nonumber
\end{eqnarray}
Where $f_s = \frac{15}{128}\frac{Gm_2a^2_1e^2}{a^3_2}$ is a scaling factor.
\subsection{Precession induced resonances} 
\label{sec:pirestheory}
In this paper we focus on precession induced resonances for which resonant angles ($\sigma_1$) assumes the following form:
\begin{equation}
	\sigma_1 = l_0f_2 + l_1\omega +l_2\Omega 
\end{equation}
where $f_2$ is the true anomaly of the companion $m_2$ and $l_0 (\ne 0),l_1$ and $l_2$ are intergers. For example, $l_0=2,l_1=-2$ and $l_3=-2$ corresponds to evection resonances.
The list of possible resonant angles is shown in Table \ref{tab:lores}. Note that we have not included resonant angles $2f_2 \pm 2\Omega$ and $2f_2 \pm \Omega$ as binaries captures in them only experience inclination variation while the eccentricity remains constant (\cite{vaillant_eviction-like_2022}).
\begin{table}
	\centering
	\caption{List of resonant angles}
	\begin{tabular}{|l|l|}
		\hline
		$2f_2-2\omega-2\Omega$ & $2f_2+2\omega+2\Omega$ \\ \hline
		$2f_2-2\omega+2\Omega$ & $2f_2+2\omega-2\Omega$ \\ \hline
		$2f_2-2\omega-\Omega$ & $2f_2+2\omega+\Omega$ \\ \hline
		$2f_2-2\omega+\Omega$  & $2f_2+2\omega-\Omega$  \\ \hline
		$2f_2-2\omega$  & $2f_2+2\omega$ \\ \hline
	\end{tabular}
	\label{tab:lores}
\end{table}

\subsection{Simplified Hamiltonian}
\label{sec:shamilt}
{ When the binary is in a certain precession induced resonance}, we can simplify the secular Hamiltonian by only keeping the resonant term of interest while ignoring other resonant terms in the expansion of $H_{m_2}$. Also, it is convenient  to use canonical transformations such that the resonant angle ($\sigma_1$) is one of the  canonical variables. Hence, we write the new canonical variables ($\sigma_0, \sigma_1,\sigma_2$) and their corresponding conjugate momenta ($\Sigma_0, \Sigma_1,\Sigma_2$) in terms of Delaunay variables \citep{murray_solar_1999}:
\begin{eqnarray}
	M&,\omega&,\Omega \nonumber \\
	L=\sqrt{G(m_0+m_1)a_1}&,J=L\sqrt{1-e^2}&,J_z=J\cos{I}  \nonumber \\
\end{eqnarray}
where $M$ is the mean anamoly of the inner orbit and $(L,J,J_z)$ are conjugate momenta of $(M,\omega,\Omega)$. Since we use the single averaged Hamiltonian, the conjugate momentum $L$ is a constant of motion which makes the semi-major axis of the binary ($a_1$) also a constant. For our canonical transformations we use, $\sigma_0=M+\omega+\Omega$, $\Sigma_0=L$,  $\sigma_2=-\Omega$ and { type-2 generating function $F_2$ of the form:} 
\begin{equation}
 F_2 = (M+\omega+\Omega)\Sigma_0 + (l_0f_2 + l_1\omega +l_2\Omega)\Sigma_1 -\Omega \Sigma_2
\end{equation}
The list of canonical variables ($\sigma_1,\sigma_2$) and their conjugate momenta ($\Sigma_1,\Sigma_2$) for various resonant angles are shown in Table \ref{tab:locv}.
\begin{table}[t]
	\centering
	\caption{List of canonical variables. }
	\begin{tabular}{|l|l|l|l|}
		\hline
		$\sigma_1$ & $\Sigma_1$ & $\Sigma_2$ & $\sigma_2$ \\ \hline
		$2 f_2-2 \omega -2 \Omega$&$\frac{L}{2}-\frac{J}{2}$&$J-J_z$&$-\Omega$ \\ \hline 
		$2 f_2+2 \omega +2 \Omega$&$\frac{J}{2}-\frac{L}{2}$&$J-J_z$&$-\Omega$\\ \hline 
		$2 f_2-2 \omega$&$\frac{L}{2}-\frac{J}{2}$&$L-J_z$&$-\Omega$\\ \hline 
		$2 f_2+2 \omega$&$\frac{J}{2}-\frac{L}{2}$&$L-J_z$&$-\Omega$\\ \hline 
		$2 f_2-2 \omega +2 \Omega$&$\frac{L}{2}-\frac{J}{2}$&$-J-J_z+2 L$&$-\Omega$\\ \hline 
		$2 f_2+2 \omega -2 \Omega$&$\frac{J}{2}-\frac{L}{2}$&$-J-J_z+2 L$&$-\Omega$\\ \hline 
		$2 f_2+2 \omega +\Omega$&$\frac{J}{2}-\frac{L}{2}$&$\frac{J}{2}-J_z+\frac{L}{2}$&$-\Omega$\\ \hline 
		$2 f_2-2 \omega +\Omega$&$\frac{L}{2}-\frac{J}{2}$&$-\frac{J}{2}-J_z+\frac{3 L}{2}$&$-\Omega$\\ \hline 
		$2 f_2+2 \omega -\Omega$&$\frac{J}{2}-\frac{L}{2}$&$-\frac{J}{2}-J_z+\frac{3 L}{2}$&$-\Omega$\\ \hline 
		$2 f_2-2 \omega -\Omega$&$\frac{L}{2}-\frac{J}{2}$&$\frac{J}{2}-J_z+\frac{L}{2}$&$-\Omega$ \\ \hline 
	\end{tabular}
	\label{tab:locv}
\end{table}
{ In addition, the Hamiltonian is modified such that $H \rightarrow H + \partial F_2/\partial t$ (\cite{goldstein_classical_1950}). Since the companion is on a circular orbit (i.e., $f_2=n_2 t$), this amounts to adding the term $l_0n_2\Sigma_1$ to the final Hamiltonian.} Putting it all together, the simplified Hamiltonian will be of the following form:
\begin{eqnarray}
	H&=&l_0n_2\Sigma_1+H_{J_2}(\Sigma_0,\Sigma_1,\Sigma_2)+ \nonumber \\ &&H_{PN}(\Sigma_0,\Sigma_1,\Sigma_2)+H_{m_2}(\sigma_1,\Sigma_0,\Sigma_1,\Sigma_2) \label{e1:hsimp}
\end{eqnarray}

\begin{figure*}
	\centering
	\includegraphics[height=0.35\linewidth,width=1.0\linewidth]{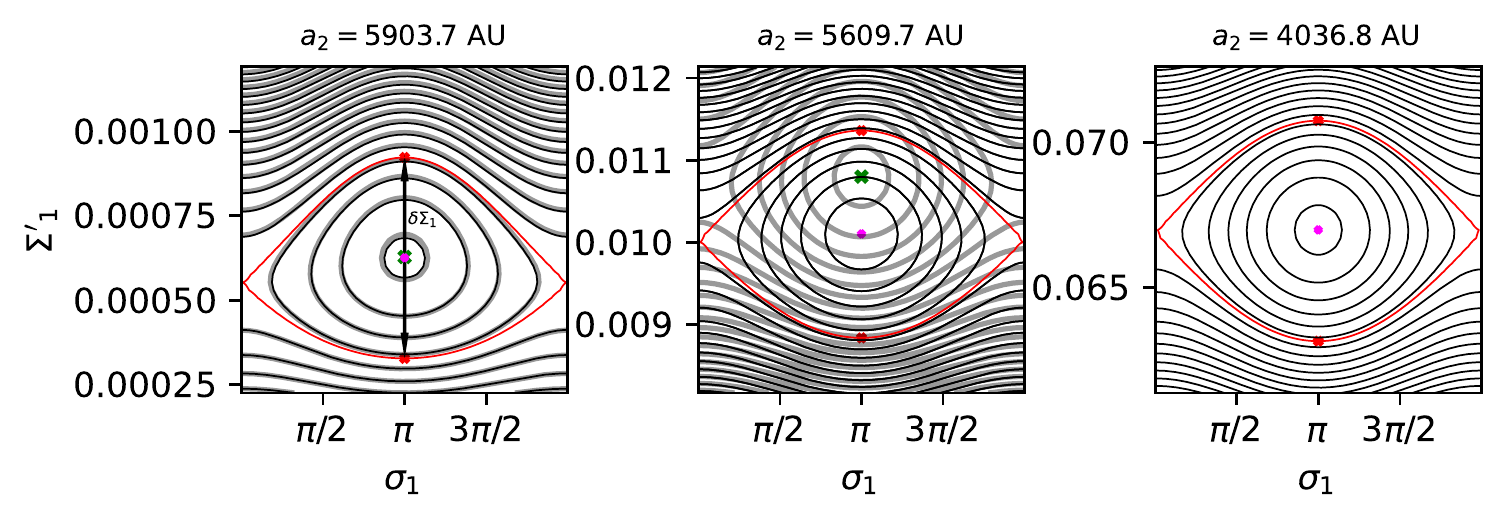}
	\caption{Contours of the simplified Hamiltonian near eviction resonance ($\sigma_1 = 2f_2-2\omega-2\Omega$). The contours of the low eccentricity Hamiltonian are shown in grey. Each panel corresponds to a different distance between the binary and the companion. The red lines show the separatrix. The extrema of $\Sigma'_1$ $(=\Sigma_1/\sqrt{G(m_0+m_1)a_1})$ on the separatrix are marked as red crosses. The fixed points of the Hamiltonian as calculated using Eqns. \ref{e1:hsimp} and \ref{eqn:hle} are marked in pink and green respectively. We can see that at low eccentricities both calculations are in agreement with each other (left panel), but they deviate at higher eccentricities (middle panel). For the calculations we choose, $m_0=500 M_\odot,m_1= 10 M_\odot,J_2R_0^2=10^{-4} $ AU$^2$ and $m_2=10^8 M_\odot$ }
	\label{fig:contours}
\end{figure*}
Note that $H_{m_2}$  can be further split into a secular term which does not depend on $\sigma_1$ and a resonant term which is proportional to $\cos{\sigma_1}$. Also, the above Hamiltonian is independent of $\sigma_2$ which makes $\Sigma_2$ a constant of motion. The analysis of the system is hence simplified by the fact that it has only one degree of freedom. 

{To illustrate the dynamical regions, we show in Figure \ref{fig:contours} the contours of the simplified Hamiltonian for eviction resonance (corresponding to $ \sigma_1 = 2f_2-2\omega-\Omega$). We choose the eviction resonance arbitrarily for illustration purposes. We can see libration regions near $\sigma_1=\pi$. The fixed points of the Hamiltonian are shown in pink. Between the panels we change the semi-major axis of the outer binary ($a_2$). We can see that as we decrease $a_2$, eccentricity of the fixed point increases. The separatrix of the Hamiltonian is shown as a red line. We define the resonance width as the difference between the maximum and the minimum of $\Sigma_1$ on the separatrix. In Figure \ref{fig:contours}, the extrema of $\Sigma'_1=\Sigma_1/\sqrt{G(m_0+m_1)a_1}$ on the separatrix are shown as red dots. }

\subsection{The location of fixed points} 
\label{sec:locfixp}
The time evolution of the system is calculated by solving Hamilton's equations for $\sigma_1$ and $\Sigma_1$: 
\begin{equation}
	\dot{\sigma_1} = \frac{\partial H}{\partial \Sigma_1}, \dot{\Sigma_1} = -\frac{\partial H}{\partial \sigma_1} \label{eq:hamilteq}
\end{equation}
In addition, the fixed points of the Hamiltonian are deduced by solving $\dot{\sigma}_1=0$ and $\dot{\Sigma}_1=0$. Since, $\dot{\Sigma} \propto \sin{\sigma}$, the fixed points occur at $\sigma=\{0,\pi\}$. {  A binary is in resonance} if it librates around $\sigma=0$ or $\pi$ which is possible only when the fixed point is stable. The other condition ($\dot{\sigma}_1 = 0$) involves complicated functions of conjugate momenta and is not trivial to solve. But usually the nodal and apsidal precession rate is dominated by the $J_2$ term and is given by: 
\begin{eqnarray}
	\dot{\omega}_{J_2}&=&\frac{3 G m_0 J_2R_0^2}{8 a_1^3 (1-e^2)^2 \sqrt{G(m_0+m_1)a_1}}(5\cos{2I}+3)	\\
	\dot{\Omega}_{J_2}&=&-\frac{12 G m_0 J_2R_0^2}{8 a_1^3 (1-e^2)^2 \sqrt{G(m_0+m_1)a_1}}\cos{I}
\end{eqnarray}
Using above equations we can write the condition for resonance as: 
\begin{equation}
	\dot{\sigma}_1 \approx l_0 n_2 + l_1\dot{\omega}_{J_2}+ l_2\dot{\Omega}_{J_2} =0 \label{eq:reslocapp}
\end{equation}

\begin{table}[h]
	\centering	
	\caption{Each row corresponds to two resonant angles. Resonant angles with negative sign after $2f_2$ are ``planar proximity angles" and the fixed points can be found for $I<I_a$ and $I>I_b$. On the other hand, resonant angles with positive sign after $2f_2$ are ``polar proximity angles" and fixed points can be found for $I_a<I<I_b$. }
	\begin{tabular}{|l|l|}
		\hline
		$\sigma_1$ & Range \\ \hline
		$2f_2\mp(2\omega+2\Omega)$ & $I_a=45.4^\circ$ and $I_b=106.9^\circ$ \\ \hline
		$2f_2\mp(2\omega+\Omega)$ & $I_a=55.0^\circ$ and $I_b=111.0^\circ$  \\ \hline
		$2f_2\mp2\omega$  & $I_a=62.4^\circ$ and $I_b=116.6^\circ$ \\ \hline
		$2f_2\mp(2\omega-\Omega)$  & $I_a=68.0^\circ$ and $I_b=123.9^\circ$  \\ \hline
		$2f_2\mp(2\omega-2\Omega)$ & $I_a=72.1^\circ$ and $I_b=133.6^\circ$  \\ \hline
	\end{tabular}
	\label{tab:lorang}
\end{table}

Note that since $n_2$ and $l_0(=2)$ are always positive, the sum of second and third terms should be negative for the fixed point to exist. For a given resonant angle this is possible only in a certain range of inclinations. This range is given in Table \ref{tab:lorang}. We classify the resonances into ``planar proximity" and ``polar proximity" resonances depending on whether or not the fixed points are possible for near coplanar configurations.

Eqn. \ref{eq:reslocapp} can be rewritten to show that the location of fixed points of the Hamiltonian are given by:
\begin{equation}
	a_2^{3} = a^3_{2,s} \frac{(1-e^2)^4}{((3+5\cos{2I})l_1-4l_2\cos{I})^2}	
\end{equation}
where $a_{2,s}$ is a scaling parameter given by:
\begin{equation}
	a^3_{2,s}=\frac{64 l^2_0 (m_0+m_1)(m_0+m_1+m_2)a_1^7}{9 m^2_0J^2_2R_0^4} \label{eq:a2loceq}
\end{equation}
Using the above expression it can be seen that for a given eccentricity, the location of fixed points of various resonances depend on the inclination of the binary. Also, for a given initial eccentricity and inclination, the order in which migrating binaries encounter various resonances depend on the term in the denominator $(3+5\cos{2I})l_1 - 4l_2\cos{I}$. Since $l_1=2$ for all the resonances, it can be seen that:
\begin{eqnarray}
	l_2 > l'_{2} \rightarrow a_2 > a'_{2}& \text{ for prograde orbits }(i<90^\circ) \nonumber \\
	l_2 > l'_{2} \rightarrow a_2 < a'_{2}& \text{ for retrograde orbits }(i>90^\circ) \nonumber
\end{eqnarray}
{  Please note that in this section, we assume that the precession of the binary is dominated by the quadrupole potential of the central object ($m_0$). We operate under this assumption throughout this paper. Also, the results derived in this section are valid even when the eccentricity of the binary is excited. This includes results on the order in which resonances are encountered and the inclination range of each resonance.}

\subsection{Low eccentricity approximation}
In this study we are mainly interested in eccentricity excitation of binaries on initially circular orbits. Consequently, in most of the systems we study the binaries are captured in the resonances at low eccentricity. Hence, in this section we derive analytical expressions for libration timescale and libration width for various resonances in low eccentricity limit. 

From Table \ref{tab:locv}, we can see that $\Sigma_1 \propto 1-\sqrt{1-e^2}$. At low eccentricities, this reduces to $\Sigma_1 \propto e^2/2$. Hence in the low eccentricity regime, the Hamiltonian can be written as a series in $\Sigma'_1=\Sigma_1/\sqrt{G(m_0+m_1)a_1}$. Keeping terms upto $\Sigma'^2_1$ in the expansion, we can write:
\begin{equation}
H_{low-e}=(c_{11}\Sigma_1+c_{21}\Sigma^2_1)+(c_{12}\Sigma_1+c_{22}\Sigma^2_1)\cos{\sigma_1} \label{eqn:hle}
\end{equation}
It can be shown that the coefficients $c_{11},c_{12},c_{21}$ and $c_{22}$ assume the following form:
\begin{eqnarray}
	c_{11} &=& 2\sqrt{G(m_0+m_1)a_1}n_2 -\frac{GJ_2R_0^2}{2a_1^3}f_{c_{11}}(\Sigma_2) \nonumber \\
	c_{12} &=& \frac{a_1^2Gm_2}{a_2^3}f_{c_{12}}(\Sigma_2,\epsilon) \nonumber \\
	c_{21} &=& -\frac{GJ_2R_0^2m_0}{a_1^3}f_{c_{21}}(\Sigma_2,\epsilon) \nonumber \\
	c_{22} &=& \frac{a_1^2Gm_2}{a_2^3}f_{c_{22}}(\Sigma_2,\epsilon)  \label{eq:coefhle}
\end{eqnarray} 
where $f_{c_{11}}$,$f_{c_{12}}$,$f_{c_{21}}$ and $f_{c_{22}}$ are function of $\Sigma_2$ (which is a constant of motion and a function of initial inclination and eccentricity) and $\epsilon$. The exact form of these functions depend on the resonant angle under consideration. Terms ${c_{12}}$ and ${c_{22}}$  represent strength of resonant perturbations from the companion $m_2$. Term ${c_{11}}$ is a measure of the closeness of the system to fixed points of the precession induced resonance. Finally, the term ${c_{21}}$ represents strength of the quadrupole potential of the binary. In this study, we assume that the precession is dominated by the $J_2$ term which means that the following relationship holds:
\begin{equation}
c_{21} >> c_{11},c_{22},c_{12}	\label{eq:coefhleap}
\end{equation}
Using the above Hamiltonian, it can be shown that the location of fixed points is given by:
\begin{equation}
	\dot{\sigma_1}=0 \implies \Sigma_1 = -\frac{c_{11}\pm c_{12}}{2(c_{21}\pm c_{22})} 
\end{equation}
In addition, the libration frequency { at exact resonance} is given by 
\begin{eqnarray}
	\lambda^2 &=& \left[ \left(\frac{\partial^2 H}{\partial \Sigma_1 \sigma_1}\right)^2-4\frac{\partial^2 H}{\partial \Sigma_1^2}\frac{\partial^2 H}{\partial \sigma_1^2} \right]_{\dot{\Sigma}=0,\dot{\sigma}=0} \nonumber \\
	&=& \pm \frac{(c_{11} \pm c_{12})(\pm c_{12}c_{22} -c_{11}c_{22} +2c_{12}c_{21})}{2(c_{21} \pm c_{22})}
	\label{eq:lamsqreq}
\end{eqnarray}
The fixed point is stable for $\lambda^2<0$ and unstable for $\lambda^2>0$. For stable fixed points, the libration timescale is given by $t_{lib} = 2\pi/\sqrt{-\lambda^2}$.

Finally, the resonant width is given by:
\begin{equation}
	\delta \Sigma_1 = \frac{\pm(\sqrt{2 c_{22} \left(c_{11}^2+c_{12}^2\right)-4c_{11}c_{12} c_{21}})}{(c_{21} \mp c_{22}) \sqrt{c_{21} \pm c_{22}}}
\end{equation}

Using Eqn. \ref{eq:coefhleap}, it can be shown that the fixed points occur at a distance $a_2$ given by:
\begin{equation}
a^3_2= \frac{a_1^7(m_0+m_1)(m_0+m_1+m_2)}{m_0^2J_2^2R_0^4}f_{a_2}(\Sigma_2) \label{eqn:locresle}
\end{equation}
It should be noted that the above expression for the location of the stable points is consistent with Eqn. \ref{eq:a2loceq} and in this limit is independent of the eccentricity.

Using Eqns. \ref{eq:coefhleap}, \ref{eq:lamsqreq} and \ref{eqn:locresle} it can be seen that the libration frequency is:
\begin{eqnarray}
	\lambda^2 &=& \frac{e^2 G m_0^3 m_2 J_2^3R_0^6}{a_1^9(m_0+m_1)^2(m_0+m_1+m_2)}f_\lambda(\Sigma_2,\epsilon) \nonumber \\
	&=& \lambda^2_s f_\lambda(\Sigma_2,\epsilon)  \label{eq:lamexple}
\end{eqnarray}
where $\lambda_s$ is the scaling parameter for the libration frequency. We can then define the scaling parameter for libration timescale $t_{lib,s} = 2\pi/\lambda_s$. This is given by:
\begin{eqnarray}
	t_{lib,s} &=& 2\pi \sqrt{\frac{a_1^9(m_0+m_1)^2(m_0+m_1+m_2)}{e^2 G m_0^3 m_2 J_2^3R_0^6}} \nonumber \\
\end{eqnarray}
$f_{a_2}$ and $f_\lambda$ are functions of $\Sigma_2$ and $\epsilon$ and their form depends on the resonant angle under consideration.
\begin{figure}
	\centering
	\includegraphics[height=0.75\linewidth,width=1\linewidth]{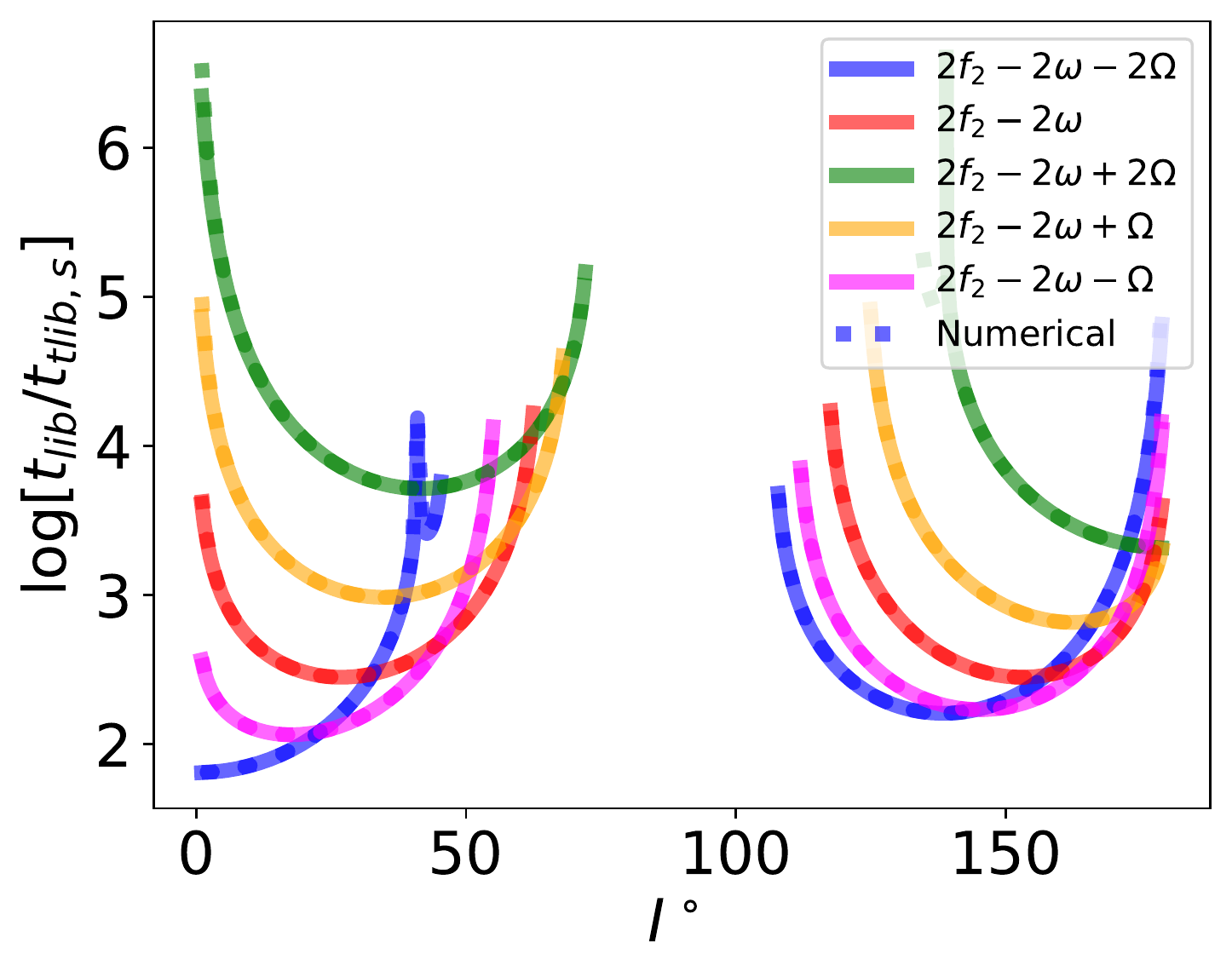}
	\caption{Libration timescale as a function of inclination of the binary for planar proximity resonances. The x axis shows inclination and the y axis shows the timescale. The solid lines show the calculation from the analytical expression Eqn. \ref{eq:lamexple}. Dotted lines show results from direct numerical calculation. We can see that they are in agreement with the analytical expression. Different colors represent different resonant angles. For binary eccentricity to be excited, libration timescale should be smaller than migration timescale. For numerical calculation we choose, $e=0.01,m_0=30 M_\odot,m_1= 30M_\odot,J_2R_0^2=10^{-3} $ AU$^2$ and $m_2=10^6 M_\odot$}
	\label{fig:libtsli}
\end{figure}

\begin{figure}
	\centering
	\includegraphics[height=0.75\linewidth,width=1\linewidth]{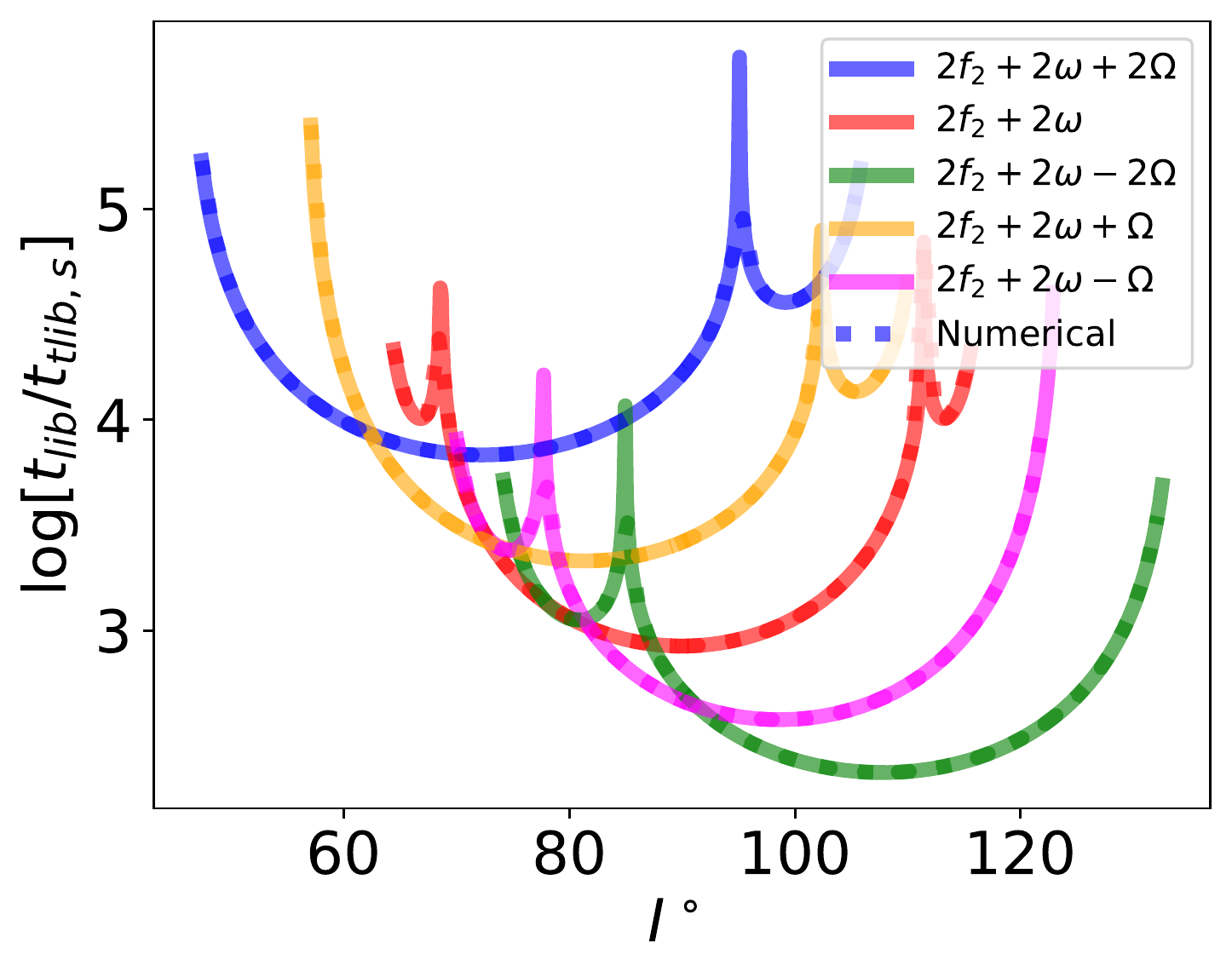}
	\caption{Same as Figure \ref{fig:libtsli} for polar proximity resonances. }
	\label{fig:libtshi}
\end{figure}
The libration timescale increases with the semi-major axis of the binary and decreases with the eccentricity, $J_2$ potential and the mass of the central object.  Figures \ref{fig:libtsli} and \ref{fig:libtshi} show the libration timescale as a function of the inclination of the binary. We can see that for a given inclination, libration timescales of different resonances can differ by orders of magnitude. At low inclination, evection ($\sigma_1 = 2f_2-2\omega-2\Omega$) and eviction ($\sigma_1 = 2f_2-2\omega-\Omega$) resonances have the shortest libration timescale. Among polar proximity resonances, those corresponding to $\sigma_1 = 2f_2+2\omega-2\Omega$ and $2f_2+2\omega-\Omega$ usually have the shortest libration timescale.  Also, the libration timescale strongly depends on the inclination of the binary. The peaks in the curves correspond to the bifurcation points. For comparison, we also show the libration timescale calculated numerically for systems whose parameters are given in the description. We can see that analytical estimates are in agreement with the numerical solutions.

\begin{table}
	\centering
	\caption{Inclination corresponding to low eccentricity bifurcation points}
	\begin{tabular}{|l|l|l|}
		\hline
		$\sigma_1$ & $I_{bif}^\circ$ & $I_{bif}^\circ$ \\ \hline
		$2f_2\mp(2\omega+2\Omega)$ & 40.98 & 95.06 \\ \hline
		$2f_2\mp(2\omega-2\Omega)$ & 84.93 & 139.02 \\ \hline
		$2f_2\mp(2\omega-\Omega)$  & 77.68 & 123.14  \\ \hline
		$2f_2\mp(2\omega+\Omega)$  & 56.86 & 102.32 \\ \hline
		$2f_2\mp2\omega$           & 68.58 & 111.42 \\ \hline
	\end{tabular}
	\label{tab:lam0inctab}
\end{table}

A bifurcation occurs when $t_{lib}=0 \implies \lambda \rightarrow \infty \implies c_{21}(\Sigma_2) \approx 0$. Taking $e=0$, one can calculate the inclination at which the above condition is satisfied. These solutions for various resonances are listed in Table \ref{tab:lam0inctab}. As we will see later, these bifurcation points determine maximum eccentricity attained as the binary trapped in the resonances. Once binaries reach these bifurcation points, they leave the resonance, which freezes their eccentricity and inclination.

Finally, the resonant width ({ see Figure \ref{fig:contours} for our definition of resonance width}) can also be similarly simplified and rewritten as:
\begin{equation}
	\delta \Sigma_1 = 2\eta \sqrt{\frac{e^2}{2\eta} -1}
\end{equation}
where $\eta = c_{12}/c_{21}$ is given by:
\begin{equation}
	\eta=\frac{45 m_0 m_2 J_2 R_0^2}{128 a_1^2 (m_0+m_1) (m_0+m_1+m_2)} f_\eta(\Sigma_2,\epsilon)
\end{equation}

\begin{figure}
	\centering
	\includegraphics[height=0.55\linewidth,width=1.0\linewidth]{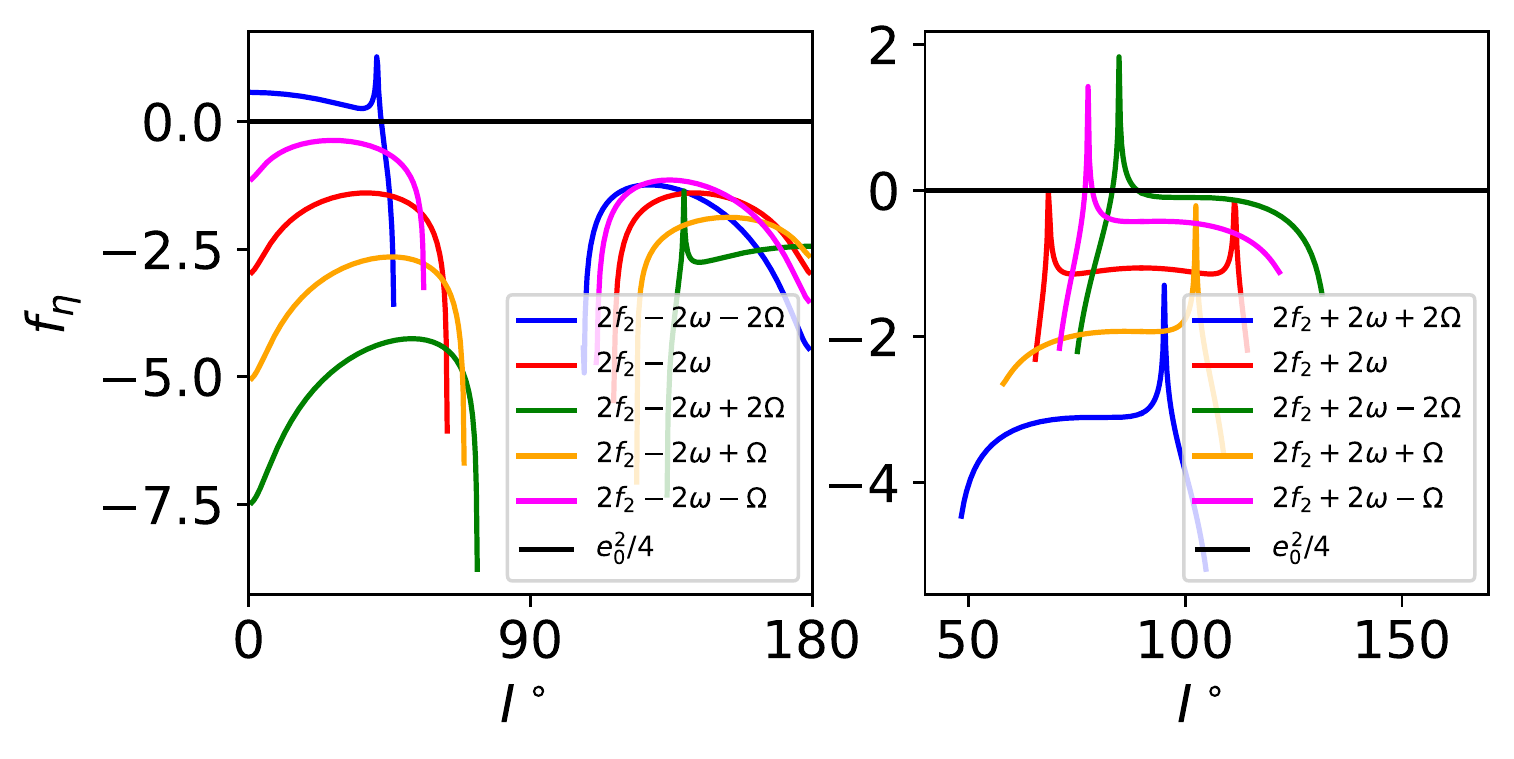}
	\caption{The function $f_\eta$ as a function of initial inclination. It is proportional to the resonance width $\delta \Sigma_1$ when smaller than $e^2/4$ (shown in solid black line).}
	\label{fig:funceta}
\end{figure}
It can be seen that the resonant width is a monotonic function of $\eta$ as long as $0<\eta<e^2/4$. Between $e^2/4$ and  $e^2/2$, the resonant width decreases with $\eta$. In the systems we focus on ($m_2 \gg m_0 >m_1$), $\eta$ mainly depends on $\Sigma_2,\epsilon,J_2R_0^2$ and $a_1$ . For $J_2R_0^2=10^{-3} $AU$^2$ and $a_1 = 0.1-1 $ AU, the factor in front of $f_\eta$ (in Eq. \ref{fig:funceta}) is the range $10^{-3}-10^{-1}$.  The specific form function $f_\eta$ takes depends on the resonance under consideration. We show $f_\eta$ as a function of inclinations $(I)$ for various resonances in Figure \ref{fig:funceta}.  Among planar proximity resonances (left panel), all but evection resonance has $f_\eta\ll e^2/4$. In addition, for all polar proximity resonances except near bifurcation points $f_\eta\ll e^2/4$. Hence, the figure can be directly used to compare resonance widths. We can see that evection and eviction resonances have the largest resonance widths. Among polar proximity resonances, $\sigma_1 = 2f_2+2\omega-2\Omega$ and $ 2f_2+2\omega-\Omega$ have the largest resonance width. Hence in the rest of the paper we focus on these resonances.

\subsection{Are secular terms important?}
{ In our discussion so far, we ignored double averaged (perturbations averaged on both the inner and outer binaries) secular effects. 
When the mutual inclination between the inner and the outer orbit is greater than $\approx 40^\circ$ double averaging secular effects von Zeipel-Lidov-Kozai (vZLK) resonance can be triggered, and it leads to large eccentricity and inclination oscillations, in addition to orbital precession (\cite{von_zeipel_1910}, \cite{kozai_secular_1962}, \cite{lidov_evolution_1962}, see \cite{naoz_eccentric_2016} for a review). 
In this section, we justify our assumption to neglect vZLK effects, since vZLK resonances are all suppressed when precession-induced resonances dominate.}

It should be noted that secular effects are suppressed when additional sources of precession (like quadrupole potential of binary components and post-Newtonian effects) dominate. We can derive the regime where secular terms are important by comparing the $J_2$ precession timescale with the vZLK timescale i.e., $\dot{\omega}_{J_2}t_k \approx 1$. Similarly, condition for evection resonance can be derived by comparing the precession timescale with the orbital period of the companion ($\dot{\omega}_{J_2} \approx n_2$). By combining the two equations, we get: 
\begin{equation}
	\frac{a_{2,kz}}{a_{2,evec}} = \frac{3J_2R_0^2m_0}{a_1^2(m_0+m_1)}	\label{eqn:evcnd}
\end{equation}
Where, $a_{2,kz}$ is the distance between the companion and the binary beyond which vZLK is suppressed and $a_{2,evec}$ is the characteristic distance where evection resonances occur. In this paper we focus on systems with $m_0 \gg m_1$ and $a_1 \in $ [0.1-1] AU. vZLK resonance is hence suppressed if $J_2R_0^2 < 10^{-2}$ AU$^2$. In the rest of this paper we keep $J_2R_0^2$ below that limit and hence ignore vZLK resonance.

\begin{figure}
	\centering
	\includegraphics[scale=0.2,height=0.8\linewidth,width=1.0\linewidth]{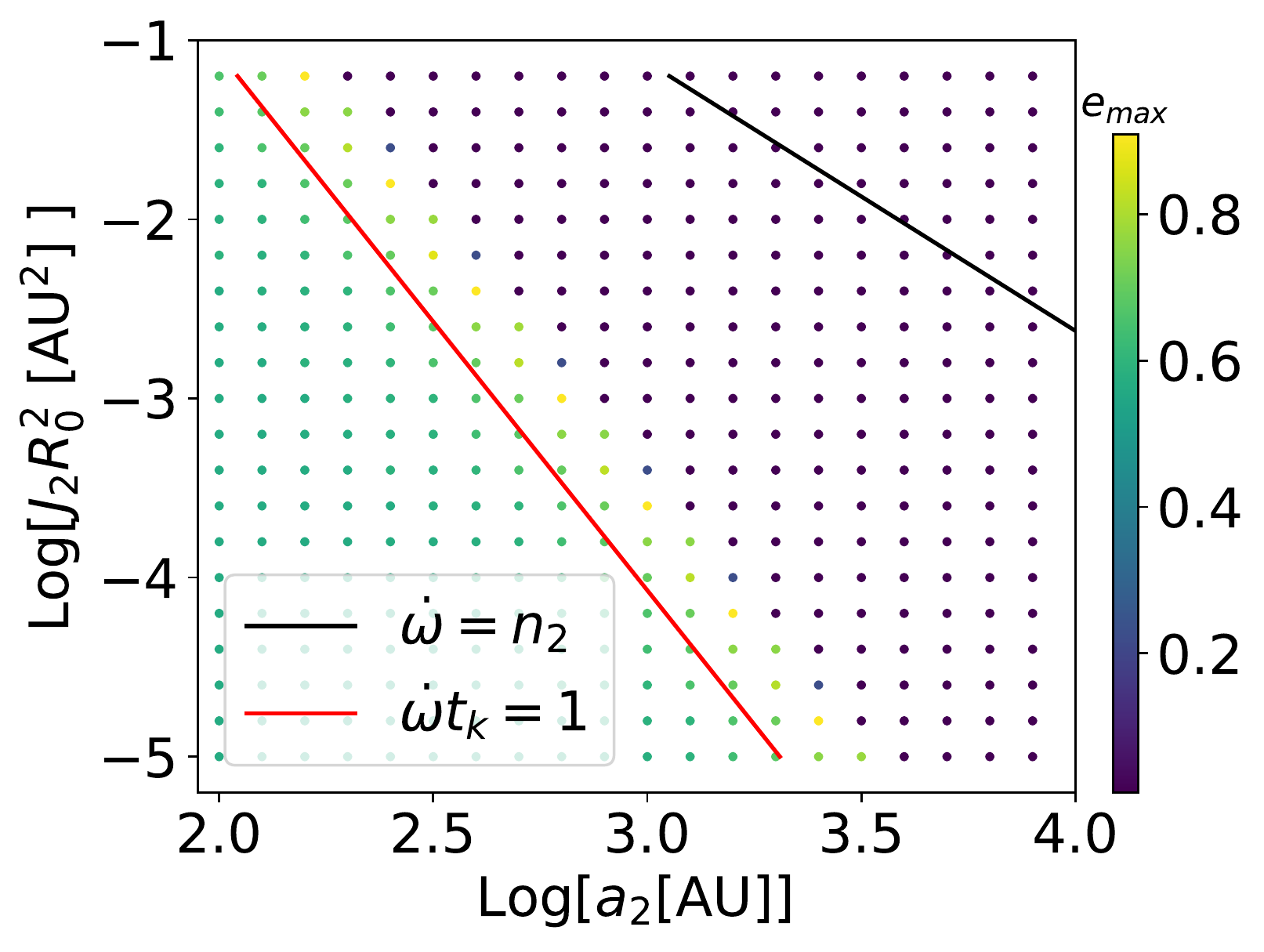}
	\caption{Outcomes of simulations with only secular terms (quadrupole order von Zeipler-Lidov-Kozai effects) of the Hamiltonian. The x axis shows the semi-major axis of the outer companion and the y axis shows strength of quadrupole potential ($J_2R_0^2$). The color corresponds to the maximum eccentricity achieved in 10 Kozai timescales($t_k$). At low values of $a_2$, the vZLK resonance dominates and the eccentricity is excited. The red line shows the semi-major axis where the precession timescale is comparable to the vZLK timescale. The black line shows the semi-major axis where precession rate is comparable to the mean motion of the outer orbit, where precession induced resonances are important. The figure shows that eccentricity excitation due to secular terms is suppressed in this regime. We use the following parameters to make this plot: $a_1=1$ AU $, m_0=10 M_\odot, m_1= 10M_\odot,\epsilon=10^\circ,I=60^\circ$ and $m_2=10^6 M_\odot$ }
	\label{fig:kzvsj2}
\end{figure}

Figure \ref{fig:kzvsj2} shows results of simulations in which we evolve the orbit of the binary using Eqn. \ref{eq:hamilteq} with a Hamiltonian containing only the secular terms ($H = H_{J_2} + H_{kz}$). For a fixed binary orbit, we change the semi-major of the companion (along x axis) and the quadrupole potential of $m_0$ (along y axis). We can see that when the binary is close to the companion, eccentricity is  excited by the vZLK resonance. At larger semi-major axes, the vZLK resonance is suppressed. Analytical estimates for characteristic length scales are shown as red and black lines. We can see that eccentricity excitation due to secular terms is suppressed where precession induced resonances are important. Also, the analytical estimates are in good agreement with the simulations. 

\section{Single averaged simulations}
\label{sec:secsim}

\begin{figure}
	\centering
	\includegraphics[height=0.8\linewidth,width=1.0\linewidth]{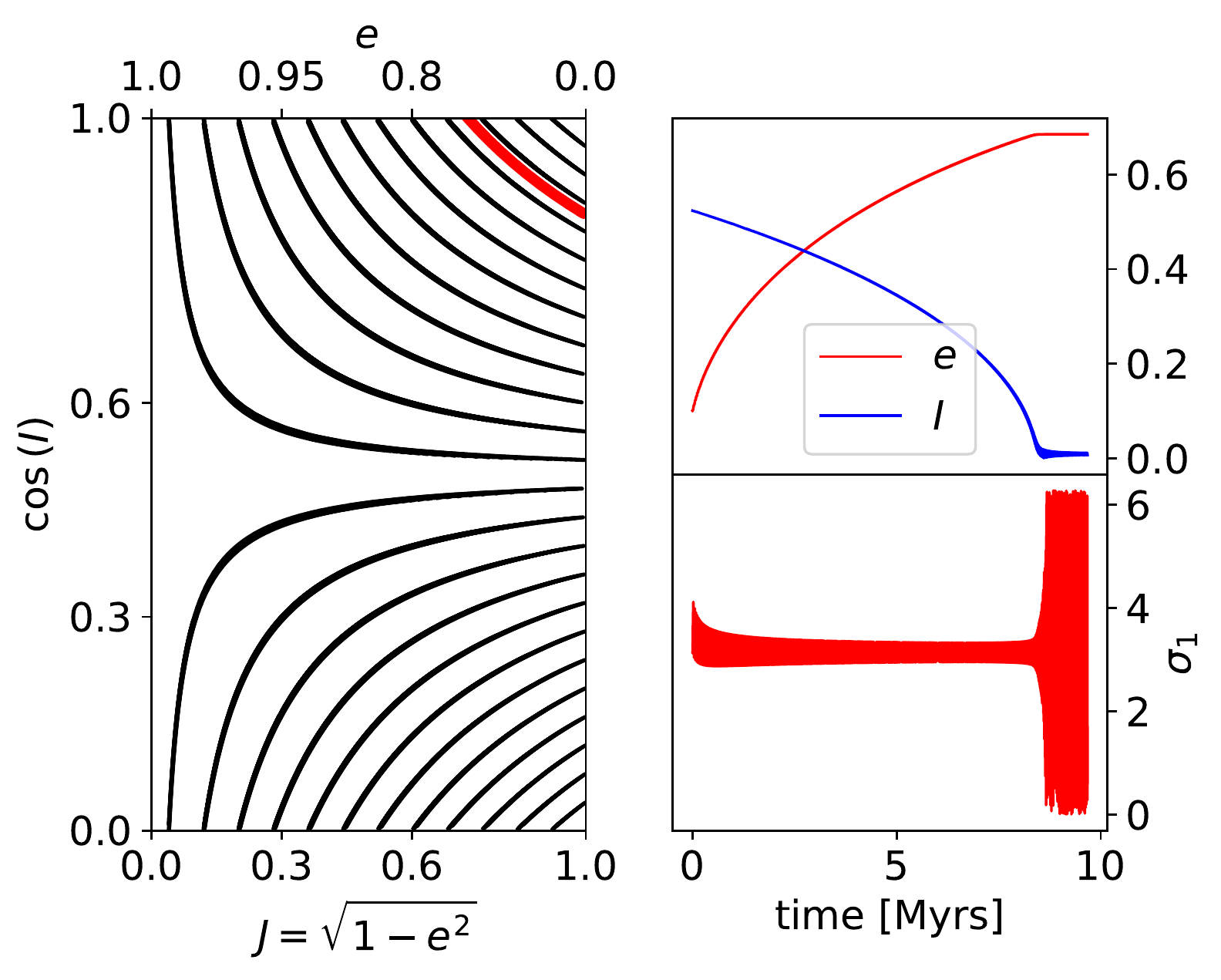}
	\caption{Eccentricity and inclination evolution for a { binary in eviction resonance} migrating towards the companion. The left panel shows contours of $\Sigma_2$, and the red line corresponds to a specific case as shown in the right panels. The right panels show the evolution of a binary starting with an initial eccentricity of 0.1 and inclination of $30^\circ$. As the binary migrates towards the companion, the eccentricity is excited to 0.6, and the inclination reduces to 0. At around 8 million years, the binary inclination reaches zero and the binary leaves resonance. At this point eccentricity and inclination is frozen. We use the following parameters to make this plot: $a_1=1$ AU$, a_2=4369.82$ AU $, \epsilon=10^\circ, m_0=30 M_\odot, m_1= 30M_\odot,J_2R_0^2=10^{-3} $ AU$^2$ and $m_2=10^6 M_\odot$ }
	\label{fig:secsim}
\end{figure}
\subsection{Single averaged code}
To study the dynamics of binaries beyond the low eccentricity limit, we numerically solve the Hamilton's equations (Eqn. \ref{eq:hamilteq}). In our simulations we use the simplified Hamiltonian as described in Section \ref{sec:shamilt} i.e. we only keep the relevant resonant term as well as the secular terms in the Hamiltonian. We use 4th order Runge-Kutta integrator from GNU scientific library to solve the equations of motion. We choose a timestep of 1/20th of libration timescale when the binary is in resonance and 1/20th of the precession timescale when it leaves the resonance.

\subsection{Eccentricity and Inclination Excitation due to Adiabatic Migration}
Once a binary is captured in a precession induced resonance, it’s eccentricity and inclination can significantly vary when the system undergoes adiabatic change. For instance, in many astrophysical systems binaries which orbit a central object are also embedded in a gas disc. Gravitational torques from the gas in the disc can cause the binary to migrate towards (or away from) the central object. This can cause $a_1$ and $a_2$ to change on timescales which depend on the local disc properties. If the migration and hardening timescales are much longer (by 1-2 orders of magnitude) than the libration timescale, the binary would stay in resonance and experience eccentricity and inclination variation. In our simulations we use the following simple prescription from \cite{lee_dynamics_2002} to model binary migration:
\begin{eqnarray}
\frac{\dot{a}_2}{a_2} = -\frac{1}{\tau_{a_2}}, \label{eqn:eqmig}.
\end{eqnarray} 
where $\tau_{a_2}$ is the migration timescale.

\begin{figure*}
	\centering
	\includegraphics[height=0.27\linewidth,width=1.0\linewidth]{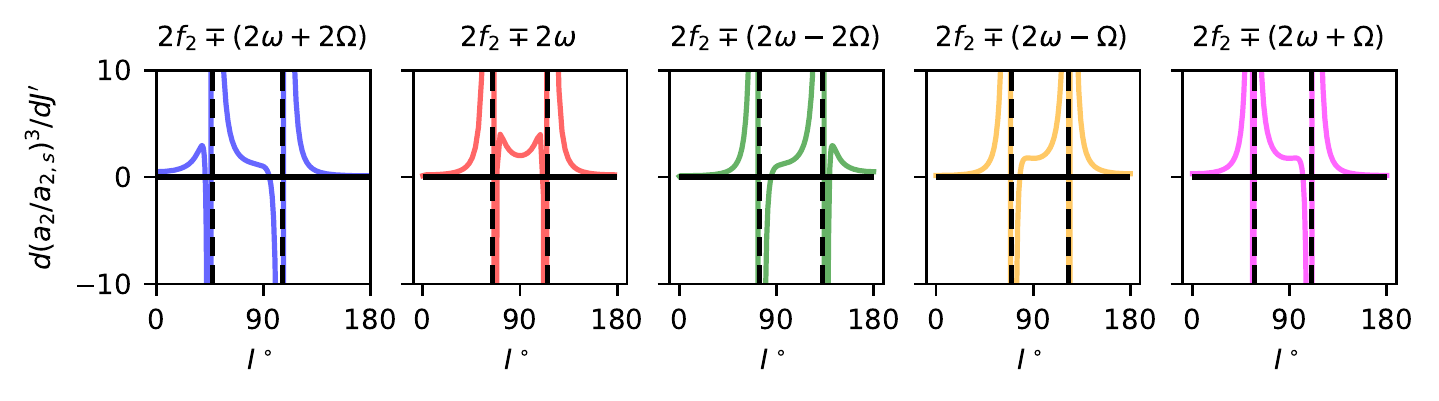}
	\caption{Regions of the parameter space where eccentricity is excited due to migration. The x axis shows the inclination of the binary and the y axis shows the derivative of $a_2/a_{2,s}$ corresponding to the location of the resonance with respect to the variable $J' = \sqrt{1-e^2}$. Each panel corresponds a different planar proximity (polar proximity) resonance as indicated by the resonant angle with -(+) sign at the top of each panel. In the regions where the derivative is positive (negative), the eccentricity increases (decreases) as $a_2$ decreases. }
	\label{fig:da2dJ}
\end{figure*}

Figure \ref{fig:secsim} shows the evolution of eccentricity (upper right panel- red curve), inclination(upper right panel-blue curve) and the resonant angle (bottom right panel-red curve) of a binary trapped in eviction resonance migrating towards the companion. The binary evolution has been calculated numerically using the method outlined above. The binary is migrating towards the companion causing the eccentricity to increase (from 0.1 to above 0.6) and the inclination to decrease ($25^\circ$ to 0). At around 8 million years, when the inclination becomes zero, the binary leaves resonance. In the bottom panel we can see that the resonant angle which was librating so far now starts circulating. After this the eccentricity and inclination are frozen. The panel on the left shows contours of $\Sigma_2$ (which is an adiabatic invariant) in black. The results from the simulations are shown in red. We can see that the binary occupies one of the contours. The functional form of  $\Sigma_2$ determines the relationship between eccentricity and inclination evolution.

It should be noted that migration towards the companion does not always excite the eccentricity of the binary trapped in a precession induced resonance. To uncover the parameter space where eccentricity is excited, Figure \ref{fig:da2dJ} shows the derivative of the scaled semi-major axis of the companion ($a_2/a_{2,s}$) with respect to the variable $J'=\sqrt{1-e^2}$ as a function of the inclination of the binary. Since $\Sigma_2$ is an adiabatic invariant, we keep it constant along these curves. We can see the derivative is positive for certain inclinations and negative for others. It crosses zero at the bifurcations points shown in Table \ref{tab:lam0inctab}. It should be noted that if the derivative is positive (negative), the eccentricity of the binary increases (decreases) when the separation between the binary and the companion decreases. For all planar proximity resonances, the derivative is positive at low inclinations. For near polar binaries, the derivative can be positive or negative depending on the inclination. In this study we focus on parameter space in which the derivative is positive.

We now look at the maximum eccentricity attained by an initially near-circular binary trapped in a precession induced resonance. We investigate eccentricity excitation using an ensemble of single-averaged simulations in which near circular binaries which are initially trapped in precession induced resonances are allowed to migrate adiabatically towards the companion. More specifically, we chose the fastest migration timescale which allows the binary to remain in the resonance.

\begin{figure}
	\centering	\includegraphics[height=0.5\linewidth,width=1.0\linewidth]{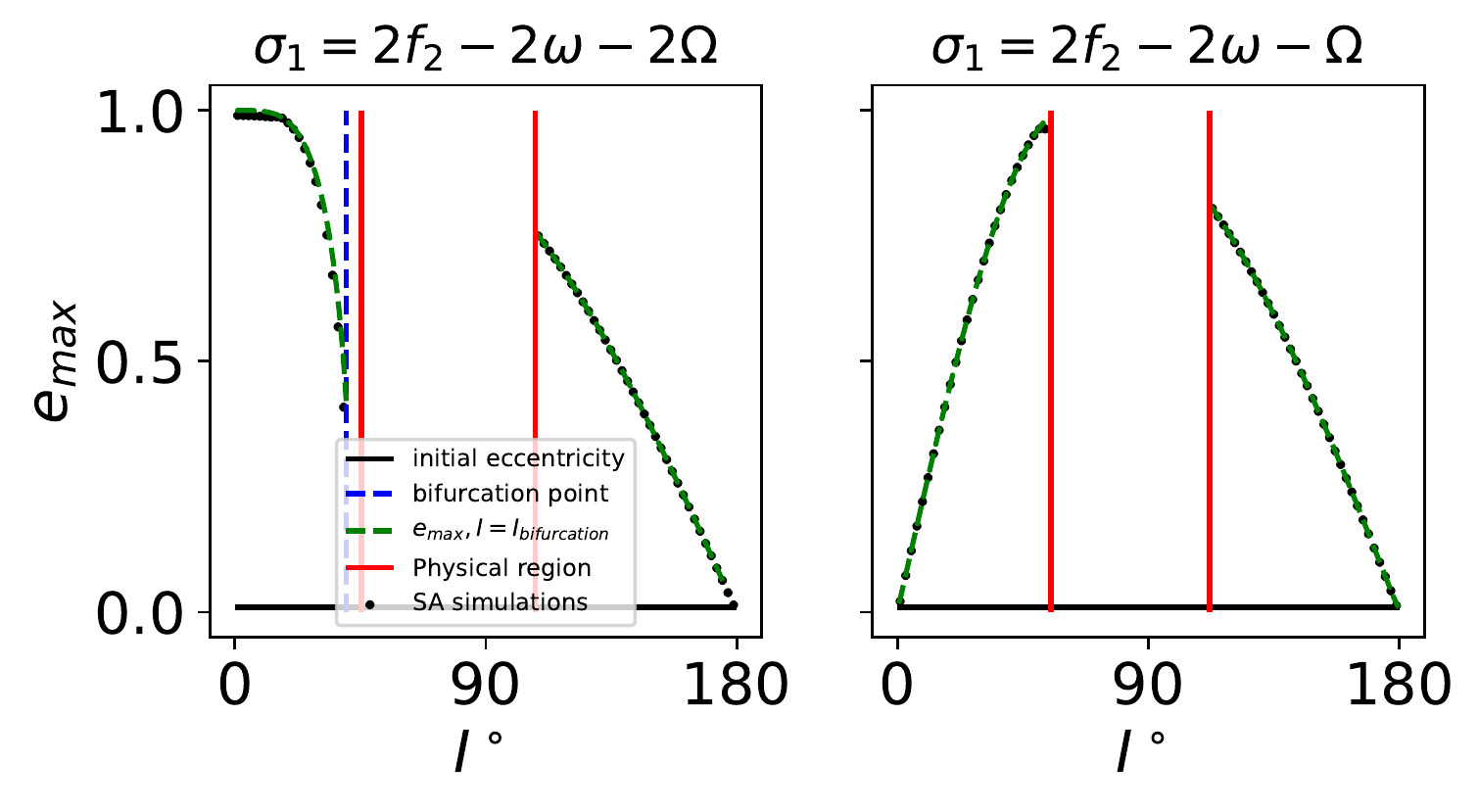}
	\caption{Maximum eccentricity as a function of initial inclination for planar proximity resonances. The x axis shows the initial inclination and the y axis shows the maximum eccentricity achieved by binaries migrating towards the companion.  The initial eccentricity is taken to be 0.2. Results from single-averaged simulations are shown as black dots. The left panel shows results for evection resonance and the right panel shows results for eviction resonance. }
	\label{fig:emaxiva2li}
\end{figure}

\begin{figure}
	\centering
	\includegraphics[height=0.5\linewidth,width=1.0\linewidth]{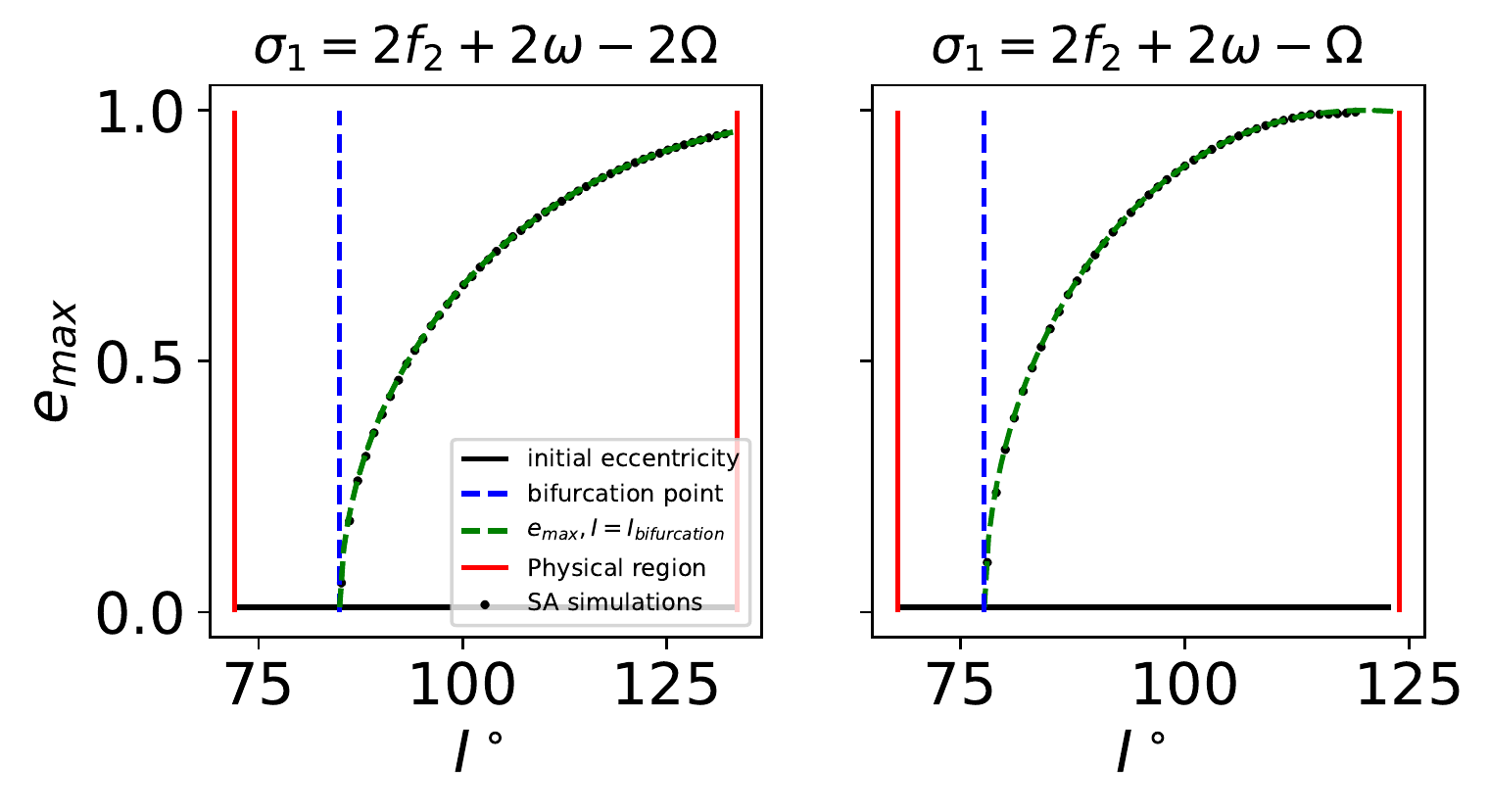}
	\caption{Maximum eccentricity as a function of initial inclination for polar proximity resonances. The x axis shows the initial inclination and the y axis shows the shows the maximum eccentricity achieved by binaries migrating towards the companion.  The initial eccentricity is taken to be 0.2. Results from single-averaged simulations are shown as black dots. The left panel shows results for $\sigma_1 = 2f_2 +2\omega-2\Omega$ and the right panel shows results for $\sigma_1 = 2f_2 +2\omega-\Omega$. }
	\label{fig:emaxiva2hi}
\end{figure}

The results of our simulations are shown in Figures \ref{fig:emaxiva2li} and \ref{fig:emaxiva2hi}. Figures \ref{fig:emaxiva2li} shows results for planar proximity resonances. It shows the initial inclination on the x axis and the maximum eccentricity attained on the y axis. The red lines show the physical region. The left panel shows results for evection resonance and the right panel for eviction resonance. The black dots show results of our single-averaged simulations. We can see that the maximum eccentricity depends on the inclination of the binary. For evection resonance, the eccentricity can reach close to unity for low inclinations ($I < 20^\circ$). This is consistent with previously studies which focus on near coplanar configurations. As the initial inclination increases, the maximum eccentricity decreases. This is due to the fact that  as the binary migrates, its eccentricity and inclination increases and at some point (when inclination $\approx 41^\circ$) the binary encounters the bifurcation point. This forces the binary to leave the resonance and further variation in eccentricity and inclination excitation of the binary is stopped.  In the region between the bifurcation point and the physical boundary for the resonance, the eccentricity of the binary decreases. This is consistent with Figure \ref{fig:da2dJ}, where we can see that at these inclinations, $\frac{da_2}{dJ'}$ is negative. For these initial inclinations, the binary eccentricity eventually reaches zero. For retrograde binaries, the migration causes the inclination to reaches $180^\circ$ at which point it leaves the resonance. 

In the figure we show analytical estimates of maximum eccentricity as green dashed lines. These are calculated by using the fact that $\Sigma_2$ is constant:
$\Sigma_2(e=0,I=I_0) = \Sigma_2(e=e_{max},I=I_{frozen})$, where $I_{frozen}$ is the inclination of the bifurcation point ($=41^\circ$, see Table \ref{tab:lam0inctab}) for $I<41^\circ$ and $180^\circ$ for retrograde orbits. It should be noted that while the estimates for the location of bifurcation points in Table \ref{tab:lam0inctab} are calculated for low eccentricity, the analytical estimate is in good agreement with single-averaged simulations.

The right panel shows the maximum eccentricity for eviction resonances. For prograde (retrograde) binaries the maximum eccentricity increases (decreases) with initial inclination. For eviction resonances, the constancy of $\Sigma_2$ dictates that as the eccentricity increases, the inclination of the binary decreases when the binary migrates towards the companion. In this case, the binaries leave the resonance when the inclination becomes 0$^\circ$ or 180$^\circ$. Hence, we choose $I_{frozen} =0^\circ$ and $180^\circ$  to calculate the analytical $e_{max}$ for prograde and retrograde binaries respectively. We can see that analytical estimates are in good agreement with single-averaged simulations.

Figure \ref{fig:emaxiva2hi}  shows results for polar proximity resonances. We again can see than the maximum eccentricity decreases for initial inclinations between the physical region boundary and the bifurcation point. For initial inclinations greater than the bifurcation point inclination, the maximum eccentricity increases with the initial inclination. In this region, as the binary migrates its eccentricity increases and inclination decreases till it encounters bifurcation point. This happens when the inclination is $\approx 85^\circ$ for the left panel  and  $\approx 78^\circ$ for the right panel. We can see that the analytical estimates for maximum eccentricity are in good agreement with single-averaged simulations.
\section{Resonance capture of Binary blackholes in AGN disks}
\label{sec:bbhagn}

\begin{figure}
	\centering
	\includegraphics[height=0.70\linewidth,width=1.0\linewidth]{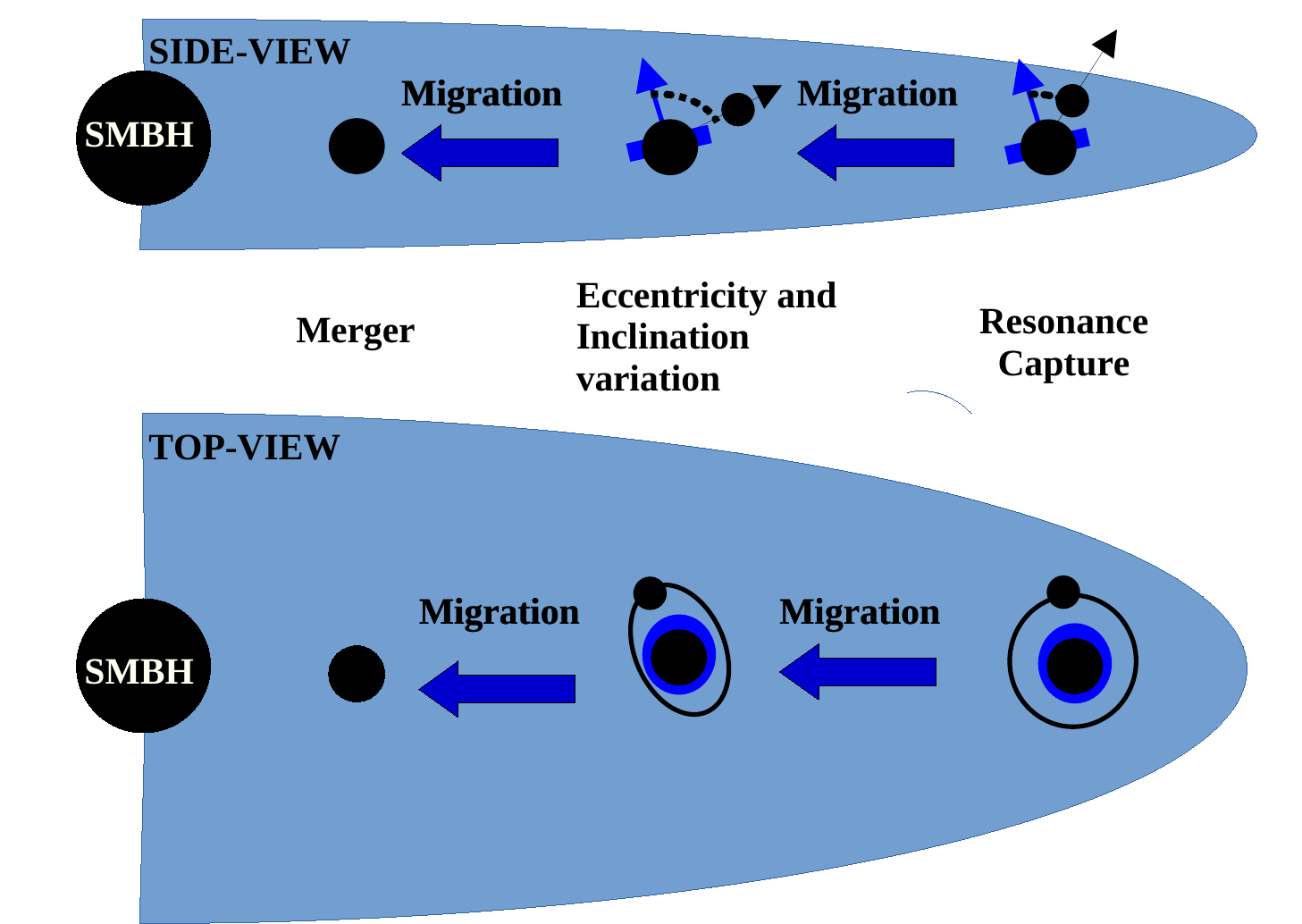}
	\caption{Binary blackholes in AGN disk. The schematic shows binary blackholes migrating in an AGN disk captured in precession induced resonance. Once captured, adiabatic migration can cause the binary eccentricity to be excited. In addition, inclination of the binary with respect to the circum-blackhole disk can also change. Eccentricity excitation can cause the binary to merge faster. }
	\label{fig:agncart}
\end{figure}

In this section we apply our results to binary blackholes in AGN disk.{  In our previous paper (\cite{bhaskar_black_2022}), we studied capture of compact binaries in evection resonance. In that work, we used a simpler setup in which the orbits of the binary and the supermassive blackhole were assumed to be coplanar. Here, we relax that assumption and allow the binary components to be misaligned with respect to the AGN disc. In the most general setup, the circum-blackhole discs of the components of the binary can have random orientations. In such a setup, the system has two degrees of freedom and the dynamics is more complicated. We leave the analysis of this general setup to a future study. Here, we focus on Intermediate mass blackhole (IMBH)-Stellar mass blackhole(SBH) binaries embedded in an AGN disc. The circum-blackhole disc of the IMBH dominates the precession of the binary orbit which allows us to neglect the precession due to the circum-blackhole disc of the SBH. This simplifies the problem and allows us to use the results derived in \S \ref{sec:antheory} and \S \ref{sec:secsim}.} 

Figure \ref{fig:agncart} shows the basic mechanism. As the binary migrates, it sweeps through precession induced resonances causing its eccentricity and inclination to change. As the eccentricity increases, the binary merger time decreases. For our fiducial simulations we focus on binary blackholes with component masses of 500 $M_\odot$ and 10 $M_\odot$. We take the disc around the 500 $M_\odot$ blackhole to have a quadrupole moment of  $J_2R_0^2 = 10^{-5}$ AU$^2$ . In addition, we assume that the disc is slightly misaligned with the angular momentum of the binary around supermassive blackhole ($\epsilon = 5^\circ$). We take the mass of the supermassive blackhole to be $10^8 M_\odot$. The binary semi-major axis is sampled between 0.1 and 1 AU.  In addition to three body dynamics described in previous sections, we also include secular changes in the orbit of the binary due to 2.5 order post-Newtonian terms. They represent orbital decay and circularization due to gravitational radiation. The corresponding averaged equations of motion are given by \cite{peters_gravitational_1964}:
\begin{eqnarray}
\frac{da_1}{dt} &=& -\frac{64}{5}\frac{G^3m_0m_1(m_0+m_1)}{c^5a_1^3(1-e^2)^{7/2}}(1+\frac{73}{24}e^2+\frac{37}{96}e^4), \nonumber \\ 
\frac{de}{dt} &=& -\frac{304}{15}e\frac{G^3m_0m_1(m_0+m_1)}{c^5a_1^4(1-e^2)^{5/2}}(1+\frac{121}{304}e^2).  \label{eqn:grrad}
\end{eqnarray}
The orbital decay timescale for a circular orbit is given by:
\begin{equation}
t_{GW} = \frac{a_1^4}{4}\frac{5}{64} \frac{c^5}{G^3m_0m_1(m_0+m_1)} \label{eqn:tmerg}.
\end{equation} 
When the binary is eccentric, the decay timescale is reduced by a factor of $(1-e^2)^{7/2}$. 
In the following subsection we discuss results of our ensemble simulations.
\subsection{Ensemble simulations}
\begin{figure*}
	\centering
	\includegraphics[height=0.45\linewidth,width=1.0\linewidth]{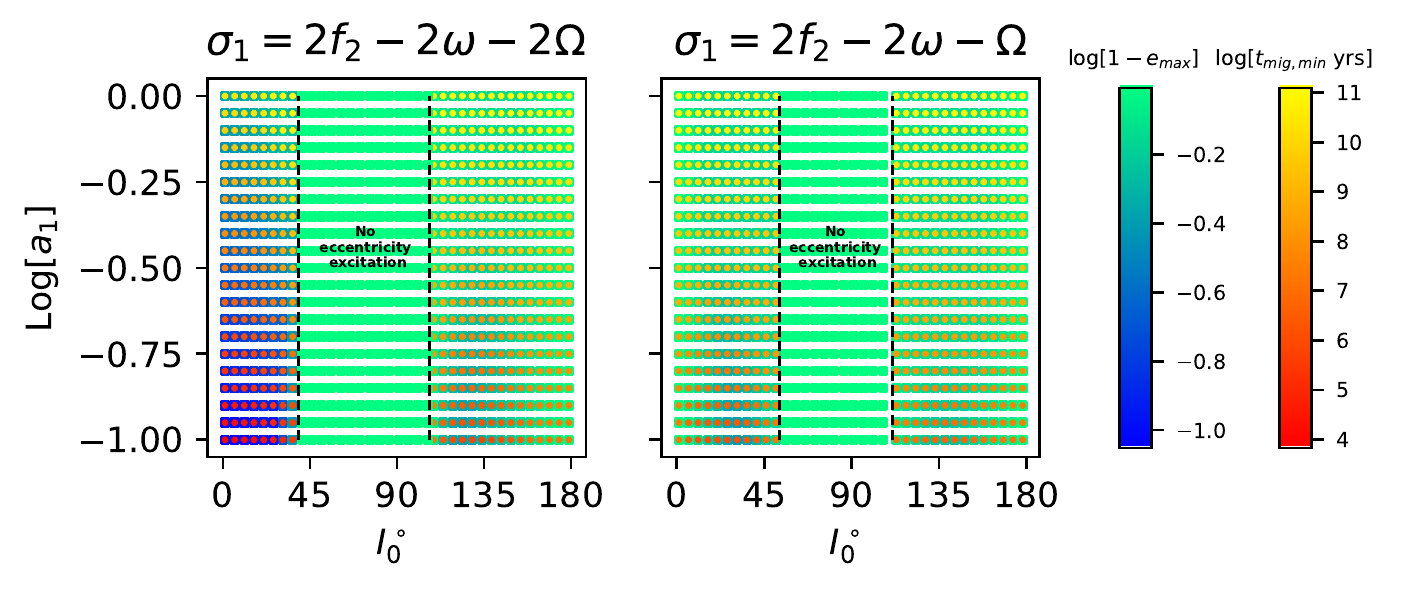}
	\caption{Results of ensemble simulations trapped in planar proximity precession induced resonances. The initial binary semi-major axis in AU is shown on the y axis and the initial inclination is shown on the x axis. Colors show the maximum eccentricity (squares) attained by binaries and the fastest migration timescale (filled circles). }
	\label{fig:enemaxli}
\end{figure*}

\begin{figure*}
	\centering
	\includegraphics[height=0.45\linewidth,width=1.0\linewidth]{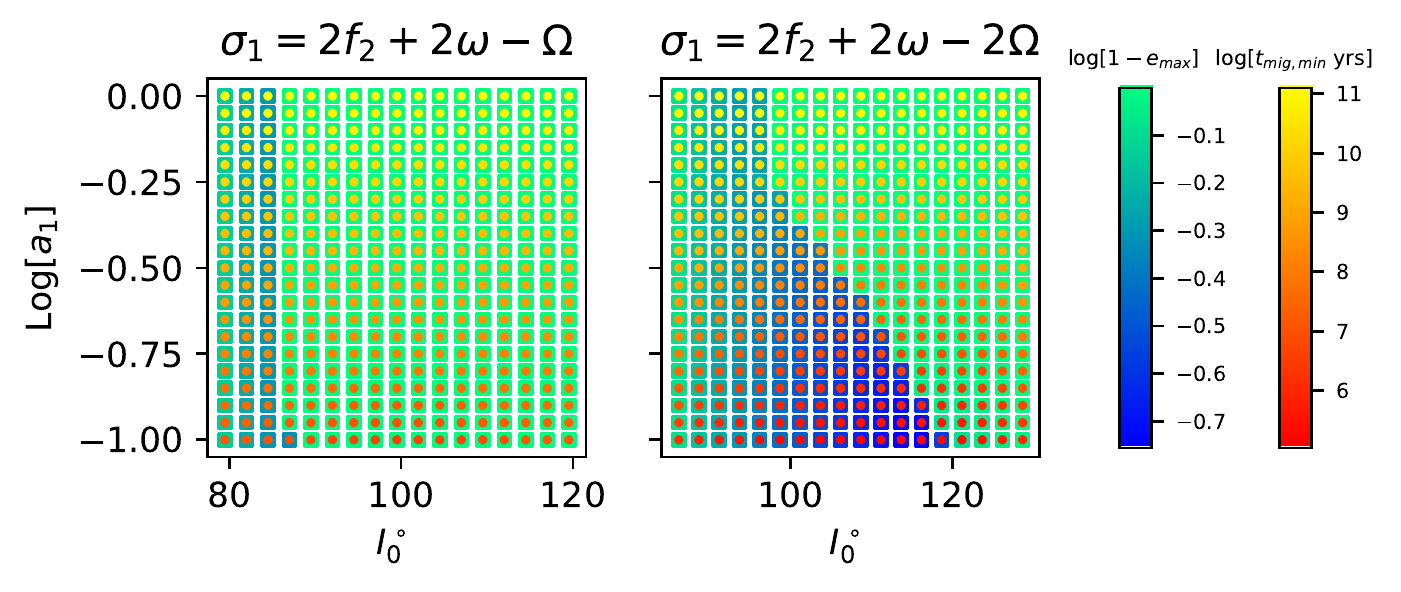}
	\caption{Same as Fig. \ref{fig:enemaxli} but for polar proximity precession induced resonances.}
	\label{fig:enemaxhi}
\end{figure*}

We run an ensemble of single-averaged simulations to study mergers of { compact binaries in precession induced resonances} in an AGN disk. We use the method outlined in Section \ref{sec:secsim} to solve equations of motion based on Eqns. \ref{eq:hamilteq}, \ref{eqn:eqmig} and \ref{eqn:grrad}.  The system parameters we use are described in the previous section.  We stop the runs when one of the following conditions is met: the binary is disrupted by the supermassive blackhole, the binary merges before it is disrupted or when the binary leaves the resonance. The binary is assumed to be disrupted when the companion is within the hill radius of the SMBH, given by:
\begin{equation}
a_{2} = 2 \left( \frac{3m_2}{m_0 + m_1} \right)^{1/3} a_1
\end{equation}
(e.g., \cite{grishin_generalized_2017}). Figures \ref{fig:enemaxli} and \ref{fig:enemaxhi} show results of our simulations for planar proximity and polar proximity resonances respectively. The x axis shows the initial inclination of the binary and the y axis shows the initial semi-major axis of the binary in AU. Colors show the maximum eccentricity (squares) and the fastest migration timescale (filled circles) which allows the binary to stay in resonance.

In Figure \ref{fig:enemaxli}, the binaries are initially trapped in evection (left panel) and eviction resonances (right panel). For evection resonance, the eccentricity is excited to large values($>0.9$) when the initial inclination is low ($i<20^\circ$). On the other hand, for eviction resonances, binary eccentricity can be excited to $0.9$ and above when the initial inclination is higher ($i \approx 40^\circ$). The maximum eccentricity can reach $1-e_{max} = 10^{-1}$. The fastest migration timescale increases with the semi-major axis of the binary. This is consistent with Eqn. \ref{eq:lamexple} which shows $t_{lib} \propto a_1^{4.5}$. The migration timescale also depends on the initial inclination. It is smallest when $I<30^\circ$ for evection resonances and when $10^\circ<I<40^\circ$ for eviction resonances.

Figure \ref{fig:enemaxhi}, shows results of single-averaged simulations for resonances corresponding to $\sigma_1=2f_2+2\omega-2\Omega$ (left panel) and $\sigma_1=2f_2+2\omega-\Omega$ (right panel). The maximum eccentricity is largest ($1-e_{max}=10^{-0.7}$) when the inclination is greater than $100^\circ$ for both resonances. The fastest migration timescale is small ($10^5$ yrs.) at high inclinations ($>100^\circ$). Please note that the results of single-averaged simulations shown in Figures \ref{fig:enemaxli} and \ref{fig:enemaxhi} are consistent with results of section \ref{sec:antheory}. { Also, the eccentricity excitation causes the binaries to merger faster, consequently enhancing the merger rate of compact binaries in AGN disc.}

\subsubsection{Location of Resonances}
\begin{figure}
	\centering
	\includegraphics[height=0.8\linewidth,width=1.0\linewidth]{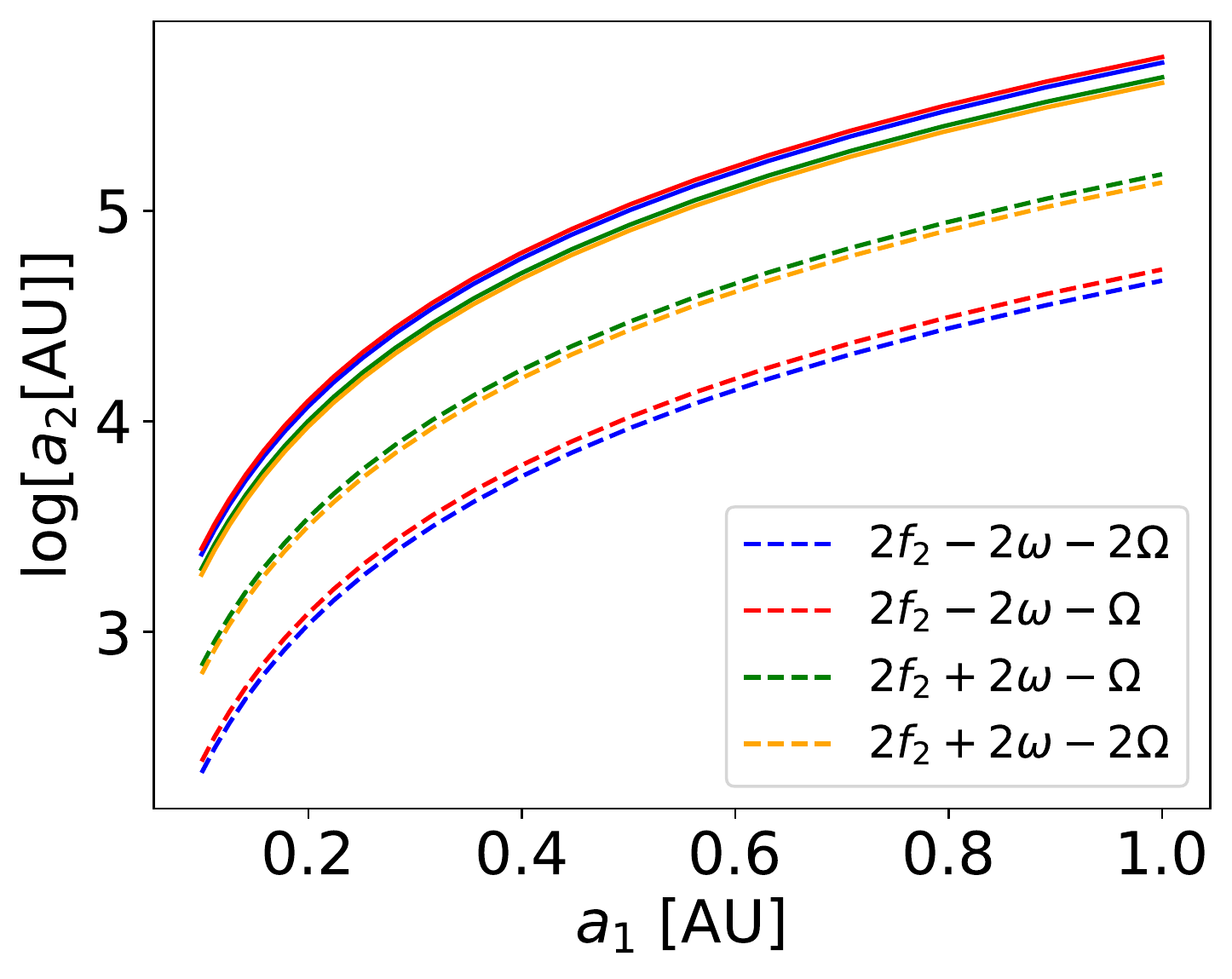}
	\caption{Location of resonances. The x axis shows the binary separation ($a_1$) and the y-axis shows the logarithm of the distance from the SMBH ($a_2$) at the location of resonances. The binaries are assumed to have an eccentricity of 0.01. Different colors correspond to different precession induced resonances.  For a given binary separation  and eccentricity, the semi-major axis of the companion depends on the inclination. The figure shows the minimum semi-major axis in dashed lines and the maximum semi-major axis in solid lines.  }
	\label{fig:a1a2}
\end{figure}

The location of resonances in the AGN disc is shown in Figure \ref{fig:a1a2}. The logarithm of the distance from the super-massive blackhole ($a_2$) is shown as a function of binary separation ($a_1$). The minimum distance from the supermassive black-hole is shown as dashed lines and the maximum separation is shown as solid lines. The spread in $a_2$ is due to the dependence on the inclination of binaries. Different colors show results for different precession induced resonances. We can see that $a_2$ increases with the binary separation. Also, for a given binary separation, $a_2$ can change by an order of magnitude depending on the inclination of the binary. Generally, the resonances occur between $10^{2.5}-10^{5.5}$ AU. A scaling relationship can be derived using Eqn. \ref{eqn:locresle}:

\begin{eqnarray}
a_2&=& 2.5 \times 10^4 \left(\frac{a_1}{0.5 \text{AU}}\right)^{7/3}\left(\frac{m_0}{500 M_\odot}\right)^{-1/3}\left(\frac{m_2}{10^8 M_\odot}\right)^{1/3} \nonumber\\
&&\times \left(\frac{J_2R_0^2}{10^{-5} \text{AU}^2}\right)^{-2/3}(f_{a_2}(\Sigma_2))^{1/3} \text{AU} \label{eqn:locreslese}
\end{eqnarray}

Please note that the factor $(f_{a_2}(\Sigma_2))^{1/3}$, which depends on the initial inclination of the binary, can vary by an order of magnitude.

\subsubsection{Maximum eccentricity}

\begin{figure}
	\centering
	\includegraphics[height=0.55\linewidth,width=1.0\linewidth]{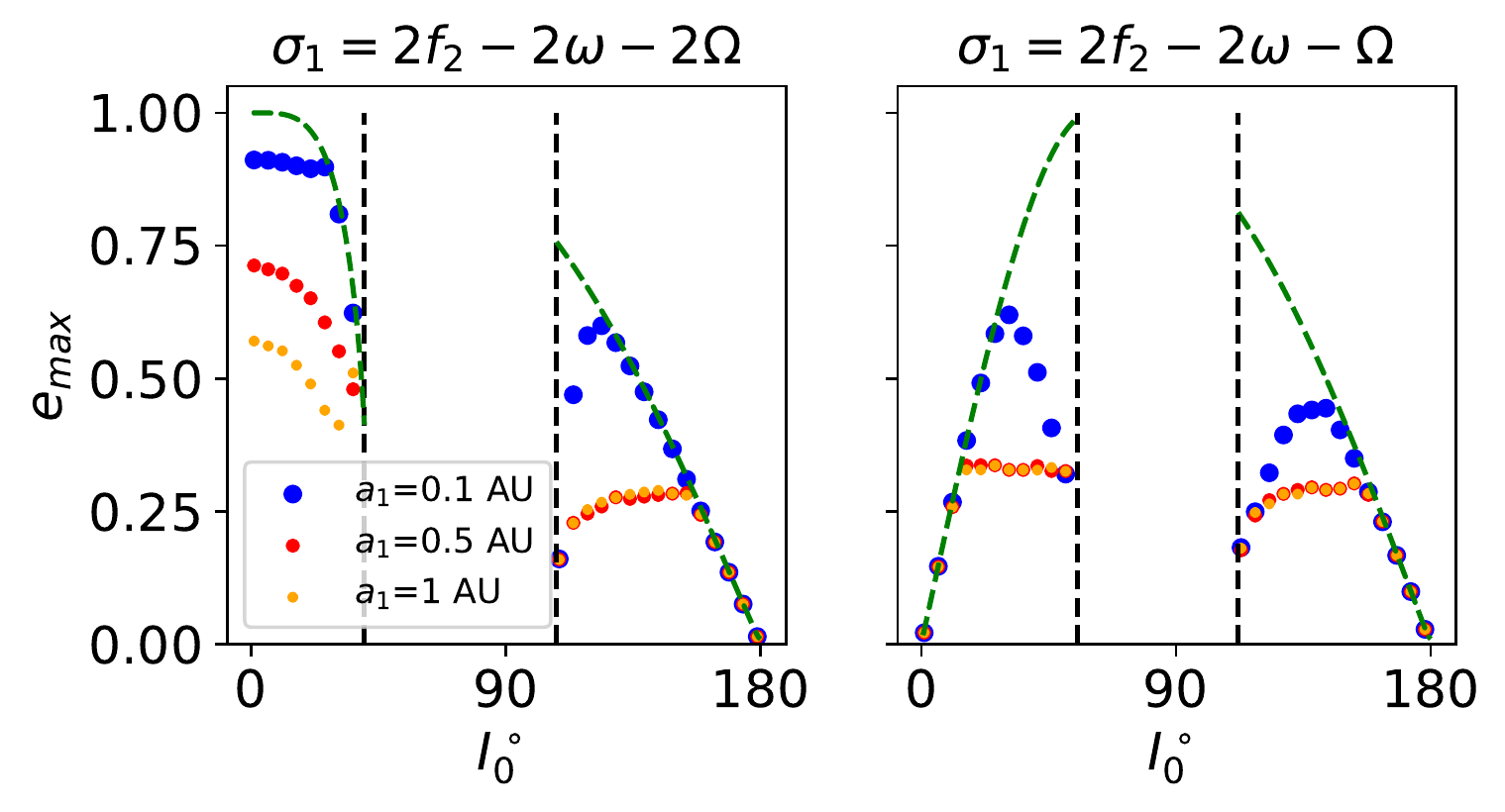}
	\caption{Maximum eccentricity achieved in our ensemble simulations. The panel on the left shows results for evection resonance and the panel on the right shows results for eviction resonance. Colored dots show results from simulations; red, blue and orange dots correspond to binaries with semi-major axis of 0.1,0.5 and 1 AU respectively. The black dashed lines show the physically allowed regions. The green dashed lines are the analytical estimates of maximum eccentricity (same as Figure \ref{fig:emaxiva2li}).  }
	\label{fig:maxeens}
\end{figure}

\begin{figure}
	\centering
	\includegraphics[height=0.55\linewidth,width=1.0\linewidth]{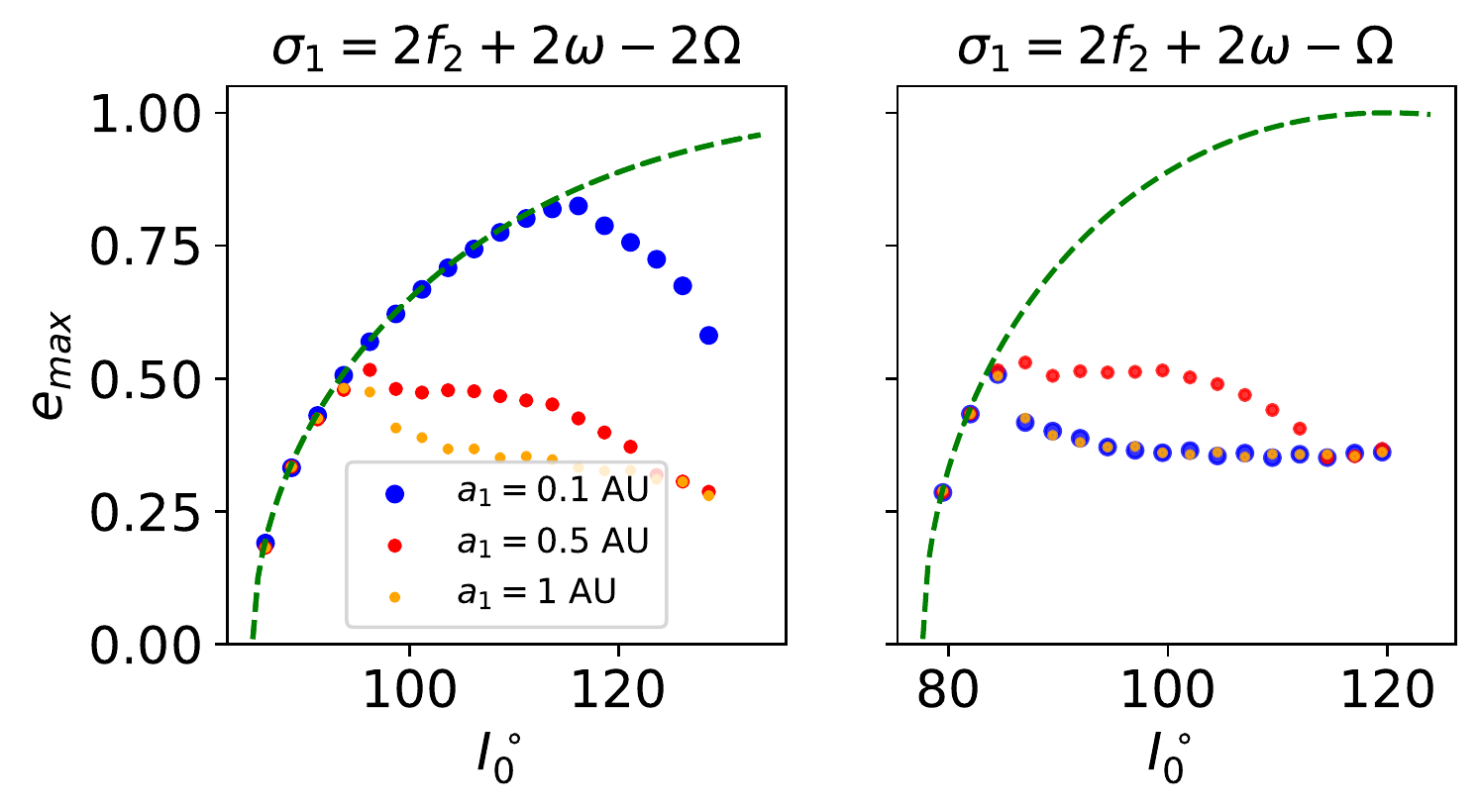}
	\caption{Same as Figure \ref{fig:maxeens} but for polar proximity resonances: $\sigma_1=2f_2+2\omega-2\Omega$ and $\sigma_1=2f_2+2\omega-\Omega$}
	\label{fig:maxeenshi}
\end{figure}

The maximum eccentricity achieved by binaries trapped in resonances as a function of their initial inclination is shown in Figures \ref{fig:maxeens} and \ref{fig:maxeenshi}. The initial binary separation is shown in different colors. The analytical estimates for maximum eccentricity (calculated without accounting for gravitational radiation) from Section \ref{sec:secsim} are shown in green. It should be noted that once a binary is trapped in a precession induced resonance, migration increases the binary eccentricity \footnote{We are only interested in parameter space where the $da_2/dJ'$ is positive (see Figure \ref{fig:da2dJ}).} and gravitational radiation reduces the eccentricity. The maximum eccentricity is reached when the following condition is met:
\begin{equation}
\left(\frac{de}{dt}\right)_{GW} = \left(\frac{de}{dt}\right)_{migration} \label{eq:ecccomp}
\end{equation}
Focusing on closely separated binaries ($a_1 \sim 0.1$ AU), we can see that the simulation results are consistent with the analytical estimates when the maximum eccentricity is low ($e_{max}<0.5$). But as the maximum eccentricity increases, gravitational radiation becomes important and suppresses the eccentricity excitation; thereby reducing the maximum eccentricity.  At larger semi-major axes, both the gravitational radiation timescale ($t_{gr} \propto a_1^{4}$) and the libration timescale ($t_{lib} \propto a_1^{4.5}$) increases. Since libration timescale (and the fastest migration timescale) increases faster than the gravitational radiation timescale, they become comparable at lower eccentricities and the eccentricity excitation is suppressed. This can be seen by looking at the data points for binaries with semi-major axis 0.5 and 1 AU (colored in red and orange). 

Finally, comparing different panels we can see that the eccentricity can be excited to 0.8 and above only when closely separated binaries are trapped in resonances corresponding to $\sigma_1=2f_2+2\omega-2\Omega$ and $\sigma_1=2f_2-2\omega-2\Omega$ (evection resonance). This is due to the fact that these resonances have the shortest libration timescales (see Figures \ref{fig:libtsli} and \ref{fig:libtshi}). This allows them to overcome eccentricity suppression from gravitational radiation. On the other hand, libration timescale is longer for resonances corresponding to $\sigma_1=2f_2+2\omega-\Omega$ and $\sigma_1=2f_2-2\omega-\Omega$ (eviction resonance) and their ecccentricity excitation is suppressed ($e_{max}<0.5$).
\subsubsection{Migration timescale}
\begin{figure}
	\centering
	\includegraphics[height=0.55\linewidth,width=1.0\linewidth]{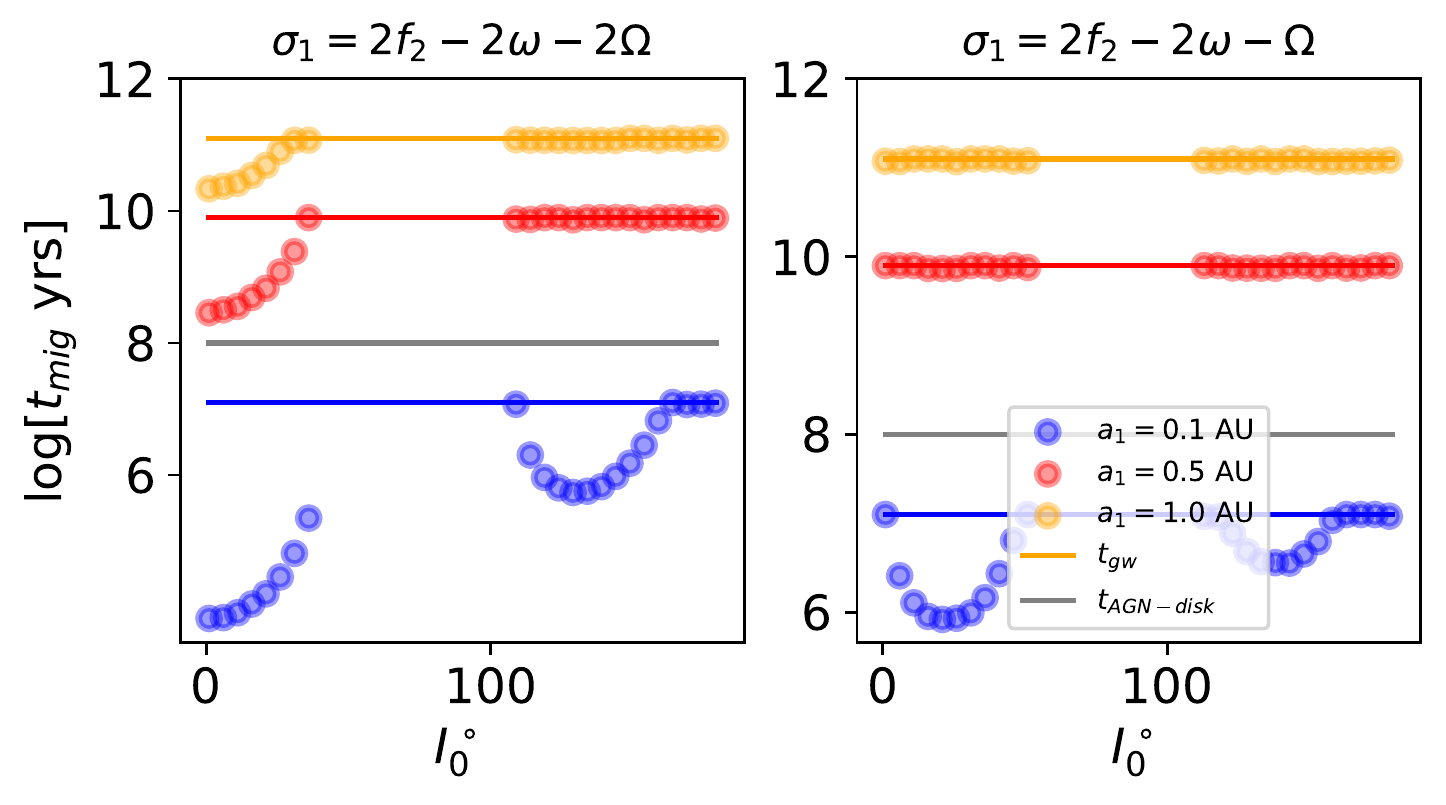}
	\caption{ Fastest migration time which allows binaries to stay in resonance. The panel on the left shows results for evection resonance and the panel on the right shows results for eviction resonance. Colored dots show results from simulations; res, blue and orange dots correspond to binaries with semi-major axis of 0.1,0.5 and 1 AU respectively. The solid lines show the gravitational radiation timescale.}
	\label{fig:tmigminli}
\end{figure}

\begin{figure}
	\centering
	\includegraphics[height=0.55\linewidth,width=1.0\linewidth]{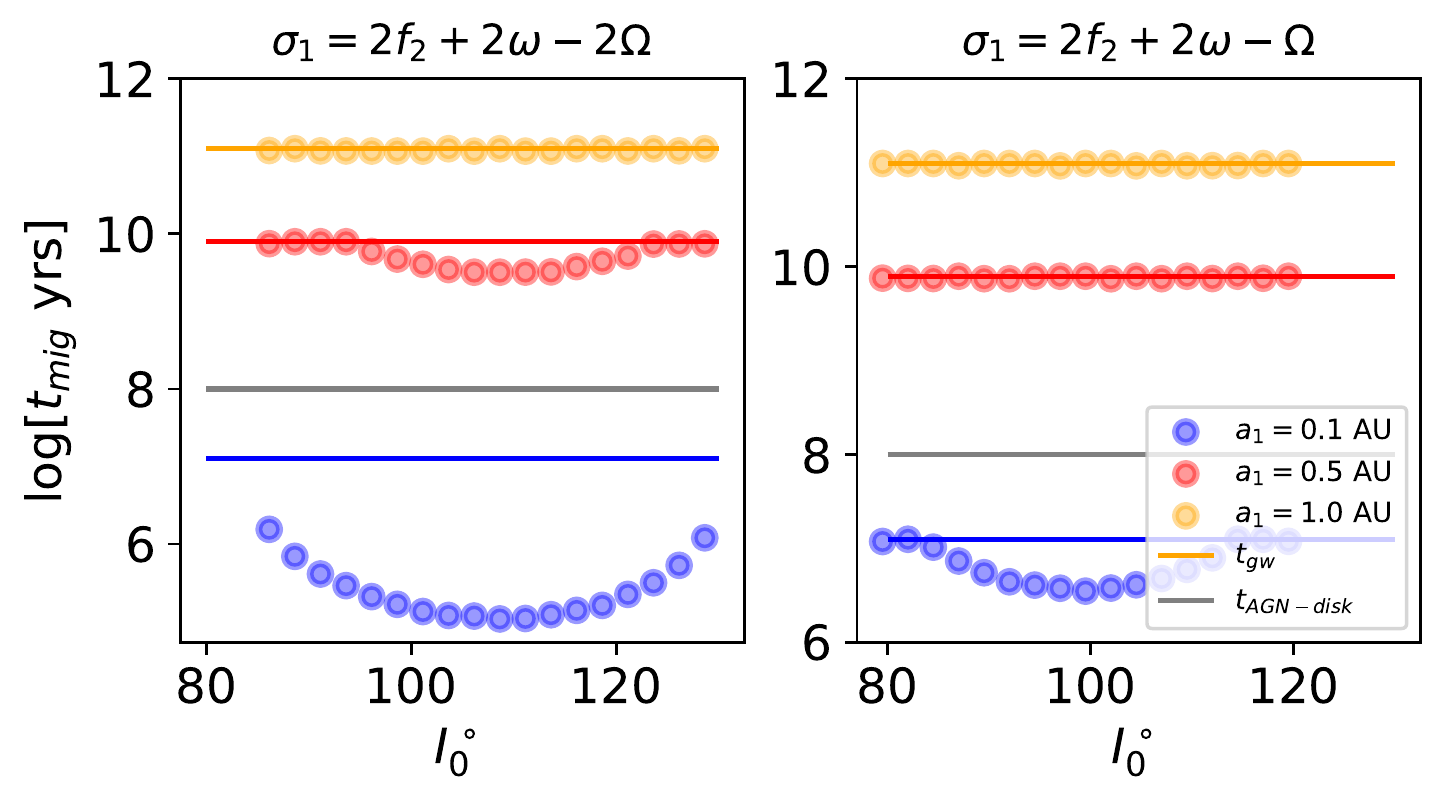}
	\caption{Same as Figure \ref{fig:tmigminli} but for polar proximity resonances: $\sigma_1=2f_2+2\omega-2\Omega$ and $\sigma_1=2f_2+2\omega-\Omega$}
	\label{fig:tmigminhi}
\end{figure}

The fastest migration timescale which keeps the binary in resonance as a function of initial inclination is shown in Figs. \ref{fig:tmigminli} and \ref{fig:tmigminhi} as filled circles. In addition, the gravitational radiation timescale is shown as solid lines. Similar to the previous plots, the semi-major axis of the binaries is shown in different colors. We can see that the migration timescale increases with the semi-major axis of the binary. The migration timescale also depends on the inclination of the binary. We can see that closely separated binaries ($a_1 \sim 0.1 $ AU) can migrate faster than the gravitational radiation timescale while keeping the binary in resonance. This allows the binary to merge on a faster timescale due to eccentricity excitation by the precession induced resonances. At larger binary separations ($a_1 > 0.5$ AU), the migration timescale is comparable to the gravitational radiation timescale. Hence in this regime, precession induced resonances are not effective in reducing the merger time. It should also be noted that the AGN disk lifetime is estimated to be around $10^{8}$ years. Hence, the parameter space where the libration timescale is longer can be ignored. In our simulations, this is true for all resonances when $a_1 \geq 0.5$ AU.

Comparing different resonances, we can see that only closely separated binaries trapped in resonances corresponding to $\sigma_1=2f_2-2\omega -2\Omega$ (evection resonance) and $\sigma_1=2f_2+2\omega -2\Omega$ can significantly excite the eccentricity. Both of them have migration timescales less than the gravitational radiation timescale and the AGN disk lifetime. The libration timescales of the other two resonances is longer and comparable to gravitational radiation timescale. Hence their eccentricity excitation is suppressed.

\subsubsection{Merger time}
\begin{figure}
	\centering
	\includegraphics[height=0.8\linewidth,width=1.0\linewidth]{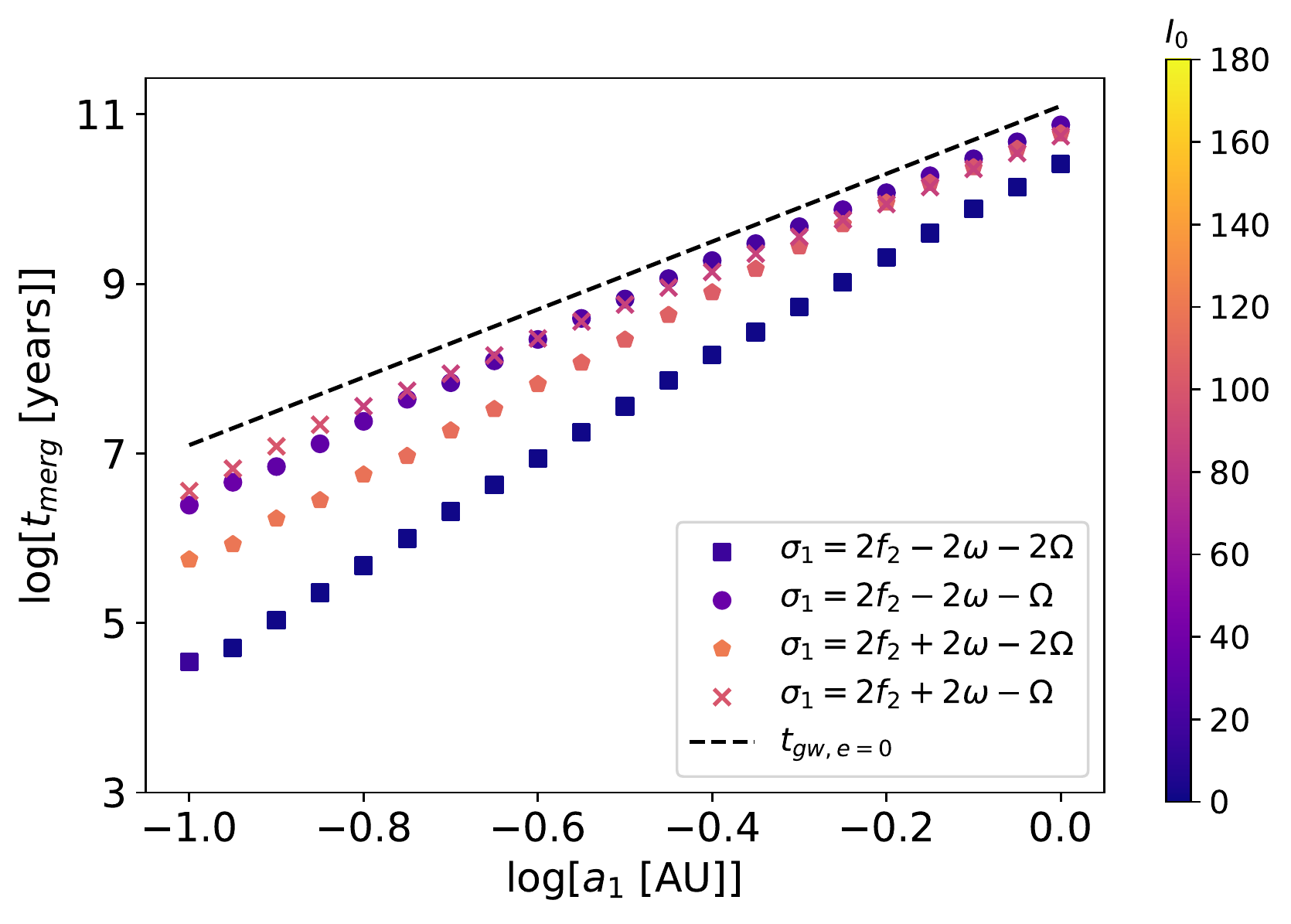}
	\caption{{ Reduced merger time due to precession induced resonances. The x axis shows the semi-major axis of the binary and the y axis shows the fastest possible merger time from our secular simulations. The color shows the initial inclination of the binary. Different markers are used to identify the resonant angles. For comparison, dashed black line shows the merger time for a circular binary whose eccentricity is not excited.}}
	\label{fig:tmerg}
\end{figure}

{ Finally, we discuss the effects of the precession-induced resonances on the SBH mergers by characterizing the merger time of migrating binaries in these resonances. Figure \ref{fig:tmerg} shows the fastest possible merger time of binaries in various precession induced resonances as a function of their semi-major axes. These results are derived from our secular simulations. We can see that the merger time increases with the semi-major axis of the binary. The merger time can be reduced by more than two orders of magnitude due to eccentricity excitation by evection resonance ($\sigma_1 = 2f_2-2\omega-2\Omega$, when $a_1 \sim 0.1$ AU). Due to longer libration timescales, binaries in resonances $\sigma_1 = 2f_2-2\omega-\Omega$ (Eviction resonance) and $\sigma_1 = 2f_2+2\omega-\Omega$ have longer merger timescales. The migration timescale is reduced by less than an order magnitude for those resonances. Binaries in $\sigma_1 = 2f_2+2\omega-2\Omega$ experience more than order of magnitude reduction in merger time. Also, by looking at the color, we can see that the initial inclination corresponding to fastest possible merger changes with the resonant angle. In addition, the dependence of the inclination on the semi-major axis of the binary is weak.  By comparing with Figures \ref{fig:libtsli} and \ref{fig:libtshi}, we can see that the inclination shown in Figure \ref{fig:tmerg} roughly corresponds to shortest libration timescale.} 

{We conclude that precession induced resonances, particularly evection resonances, can significantly reduce the merger timescale of the binaries, and hence contribute to enhanced blackhole mergers in AGN disc.}

\subsection{Resonance Capture Probability}
\label{sec:rcprob}
{  In our simulations so far we initialized the binaries at the location of resonances. But binaries crossing a resonance will not always be captured in it. Hence, we calculate the resonance capture probability of various precession induced resonances by running an ensemble of N-body simulations. Please note that this is different from our previous paper where we used secular simulations to calculate capture probability. As N-body simulations track the dynamics in greater detail, the capture probabilities we calculate here are more accurate. For these simulations, we use the \texttt{Rebound} package \citep{rein_and_liu_2012}. In our code we add extra forces to simulate precession due to quadrupole moment of the circum-blackhole discs ($J_2$ precession), post-Newtonian precession (1 PN order) and gravitational radiation (2.5 PN order) (e.g., \cite{blanchet_2006,gultekin_three_2006}).}

Binary orbits are initialized with an eccentricity of $0.01$ and their initial inclination is sampled between 0$^\circ$ and 180$^\circ$. Initial separation of the binary from the supermassive blackhole ($a_2$) is chosen such that the binary starts outside the location of fixed points of the resonance (as given by Eqn. \ref{eq:a2loceq}). The binaries are then allowed to migrate towards the supermassive blackhole on a timescale of $10^6$ years. The binary semi-major axis is chosen to be 0.5 AU. For a given initial inclination, we run 100 simulations in which resonant angle($\sigma_1$) is chosen uniformly between 0 and 360$^\circ$. The capture probability is calculated by counting the fraction of runs in which the binary is captured in resonance and it's eccentricity is excited. 

\begin{figure}
	\centering
	\includegraphics[height=0.55\linewidth,width=1.0\linewidth]{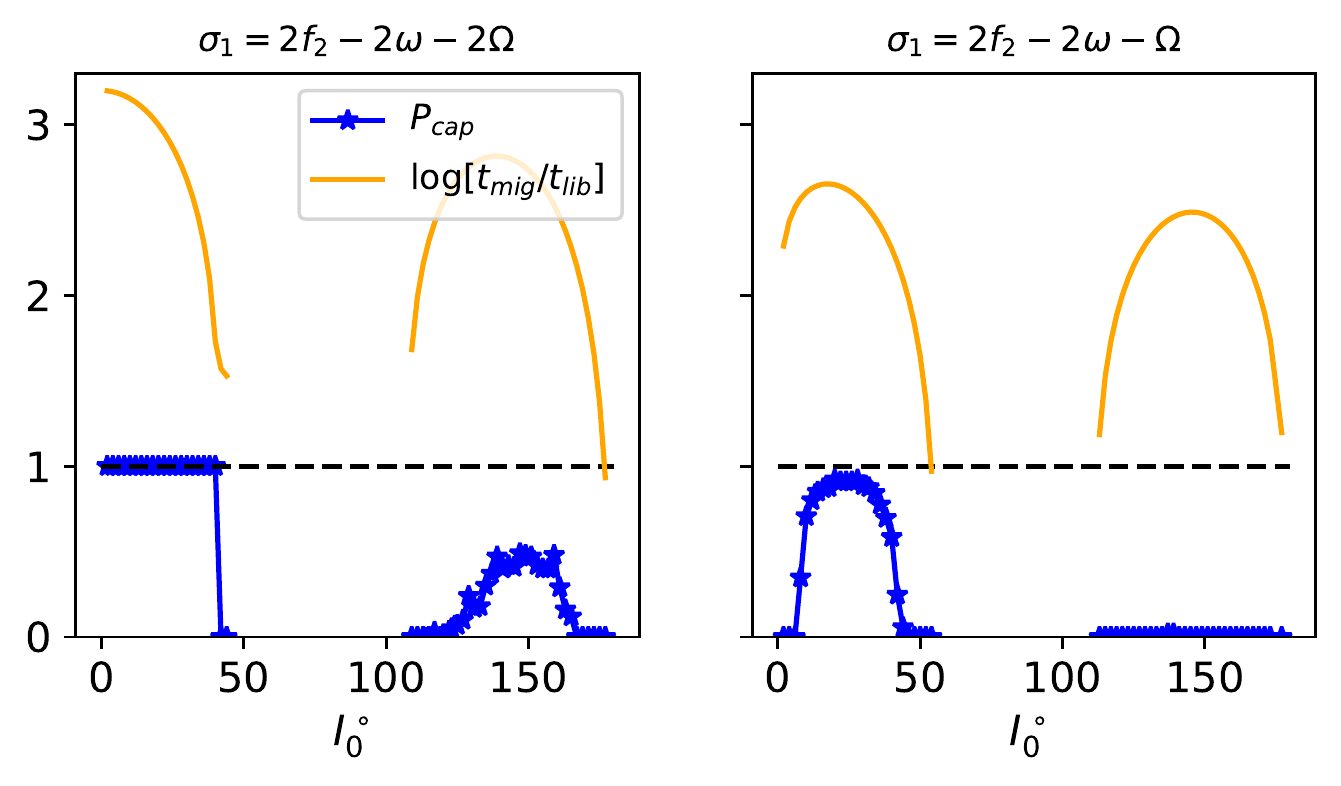}
	\caption{The capture probability of evection and eviction resonances from N-body simulations (blue). For comparison, the ratio of migration timescale with respect to libration timescale is also shown (orange). We can see that capture probability depends on the inclination of the binary. For evection resonances, capture probability is non-zero at low inclination and near $\sim 150^\circ$. On the other hand, capture probability for eviction resonance peaks at $40^\circ$ and $140^\circ$. We can also see that capture probability is high when the ratio of migration timescale with respect to libration timescale is large. }
	\label{fig:rescaptprobli}
\end{figure}

\begin{figure}
	\centering
	\includegraphics[height=0.55\linewidth,width=1.0\linewidth]{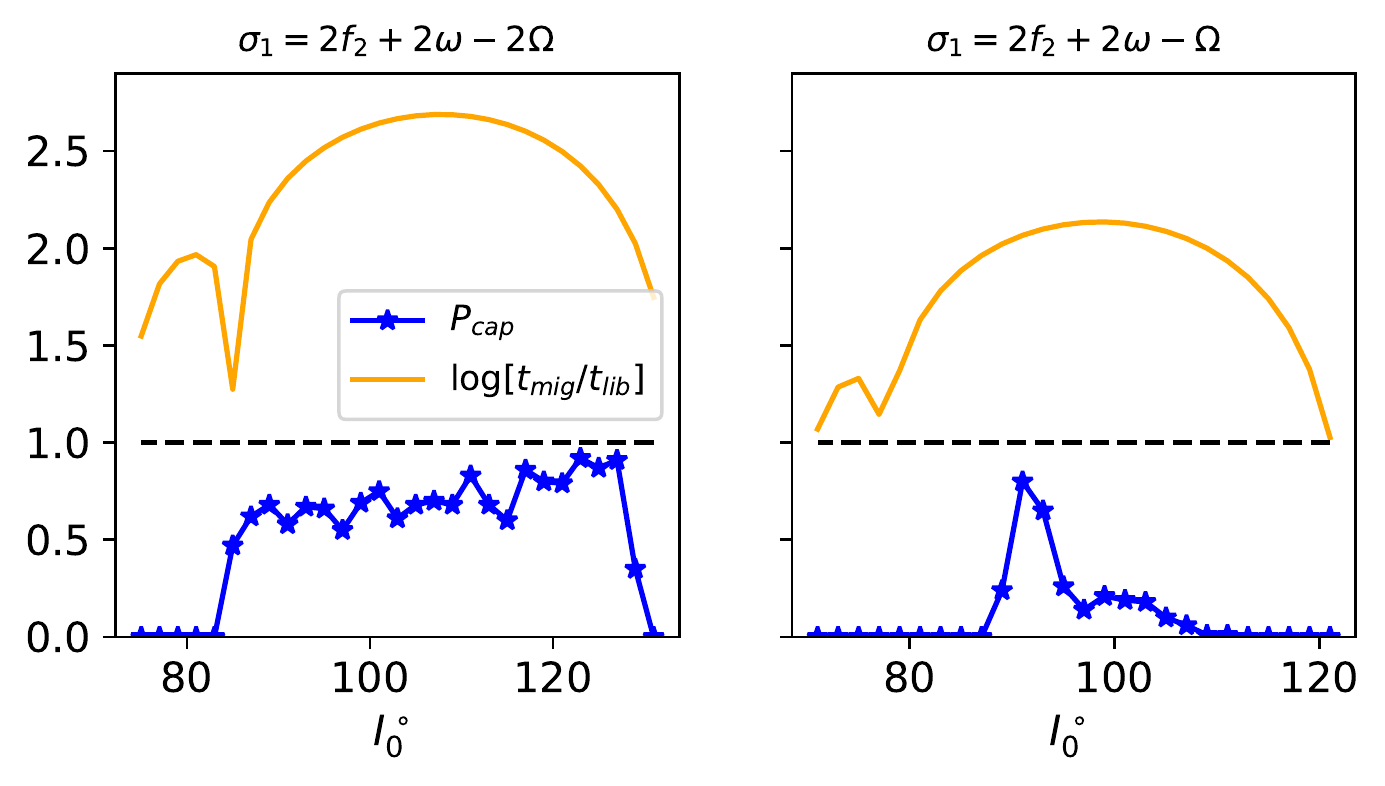}
	\caption{ Same as Figure \ref{fig:rescaptprobli} but for polar proximity resonances. For $\sigma_1 = 2f_2+2\omega-2\Omega$, capture probability is non-zero for $85^\circ<I<130^\circ$. On the other hand, for $\sigma_1 = 2f_2+2\omega-\Omega$, capture probability is non-zero for $85^\circ < I < 110^\circ$ }
	\label{fig:rescaptprobhi}
\end{figure}

The results from these simulations for evection and other resonances are shown in Figures \ref{fig:rescaptprobli} and \ref{fig:rescaptprobhi}. For comparison we also plot the ratio of migration timescale with respect to the libration timescale in orange. We can see that the capture probability depends on the initial inclination of the binary.  It is unity when the inclination of binaries captured in evection resonances is $<40^\circ$. For retrograde orbits the capture probability for evection resonances peaks at $150^\circ$. For eviction resonance, the probability is high when $10^\circ<I<40^\circ$. In addition, the capture probability is high ($>0.5$) when $85^\circ<I<125^\circ$ for $\sigma_1=2f_2+2\omega-2\Omega$ and $I\sim 90^\circ$ for $\sigma_1=2f_2+2\omega-\Omega$. Comparing with the orange curves, we can see that the capture probability is high when the migration timescale is slow compared to the libration timescale.

\subsection{Multiple resonance capture and resonance location}
In this study so far, we have focused on circular binaries captured in one of the precession induced resonances. But it is also possible for the migrating binary to be captured in multiple resonances one after another. Focusing on planar proximity resonances first, we show in section  \ref{sec:locfixp} that a migrating binary first encounters evection resonance, followed by eviction resonance and other resonances with low width. Evection resonances have large resonance width, and the binaries can hence be captured in evection resonances with high probability (see section \ref{sec:rcprob}). Once captured, the eccentricity and the inclination of the binary would increase as it migrates towards the companion. At some point the binary  would  encounter the bifurcation and leave the resonance (see Figure \ref{fig:emaxiva2li}). The binary eccentricity and inclination are frozen at this point as it migrates further towards the companion. Since the order given in section  \ref{sec:locfixp} is valid for all eccentricities, the binary would eventually encounter the eviction resonance. If captured in the eviction resonance, the binary eccentricity can be further increased at the expense of inclination. Capture into other resonances is also possible.  

\begin{figure}
	\centering
	\includegraphics[height=0.8\linewidth,width=1.0\linewidth]{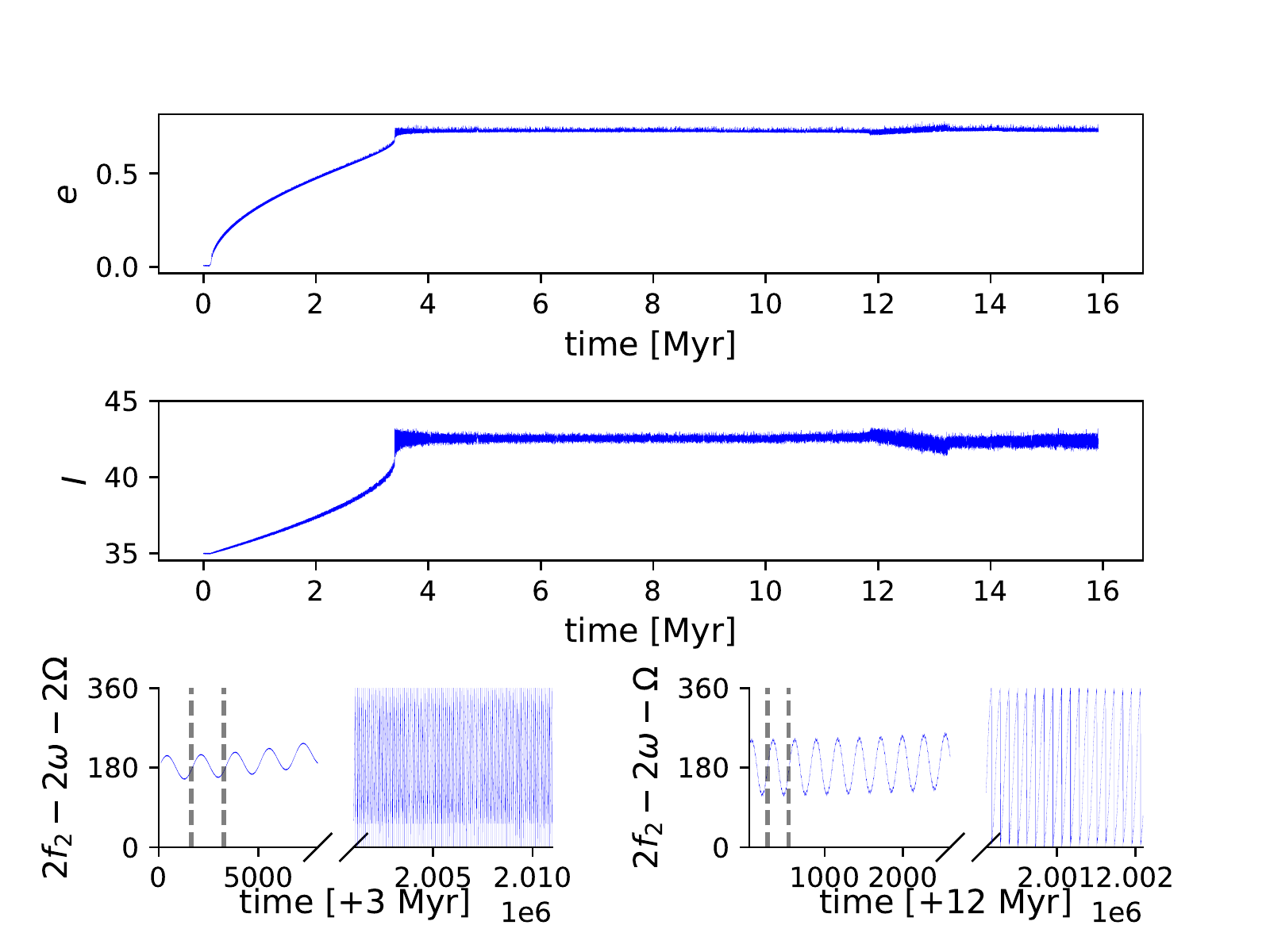}
	\caption{Capturing binary in multiple resonances (based on N-body results). The figure shows the time evolution of eccentricity ($e$), inclination ($I$) and resonant angles ($\sigma_1$) for evection (left) and eviction (right) resonances. The initial conditions are chosen such that the binary is initially captured in evection resonance. Additionally, the binary is migrating towards the supermassive blackhole causing the eccenticity and inclination to increase. At around $3\times 10^6$ years, inclination reached $40^\circ$ and the binary leaves evection resonance. As the binary keeps migrating it encounters other eviction-like resonance at $1.2 \times 10^7$ causing further eccentricity and inclination variation. { The bottom panels show the resonance angles for evection (left) and eviction (right) resonances. The dashed black lines show the libration timescales.} }
	\label{fig:mulcapres}
\end{figure}

The likelihood of capture into multiple resonances can be determined by the product of capture probability of each capture. While an extensive study of multiple captures is not in the scope of this work, we show an example in Figure \ref{fig:mulcapres}. We can see that the eccentricity of the binary initially captured in evection resonance increases to 0.72 and the inclination reaches $\approx 42^\circ$. The binary then leaves the evection resonance. Till 12 million years, the binary keeps migrating inward  when it encounters eviction resonance. This causes the inclination to decrease and eccentricity to increase. {  The resonance capture can be confirmed by looking at the bottom two panels which show the resonant angle for evection (left) and eviction (right) resonances. We can see that the resonance angle for evection resonance librates when it is in resonance at around 3 million years, and circulates at 5 million years as it is out of resonance. Similarly, at around 12 million years, the resonant angle for eviction resonance librates, but later at 14 million years it circulates.}

\section{Conclusions}
\label{sec:conc}
In this paper we study the dynamics of precession induced resonances. We identify 10 such resonances in the quadrupole expansion of the Hamiltonian. Binaries trapped in 8 of these resonances can experience eccentricity excitation. We classify these resonances into planar proximity and polar proximity resonances depending on whether or not their fixed points occur in coplanar configuration. We derive analytical expressions for location of fixed points, resonance width and libration frequency for these resonances at low eccentricity. We find that many of these resonances have small resonance width and are less likely to be captured in resonances. Hence we focus on 4 resonances: $\sigma_1 = 2f_2-2\omega-2\Omega$ (Evection resonance), $2f_2-2\omega-\Omega$ (Eviction resonance), $2f_2+2\omega-2\Omega$ and $2f_2+2\omega-\Omega$, which have wider resonance widths and shorter libration timescale. We find that the maximum eccentricity that an initially near-circular binary can achieve depends on it's initial inclination. The binary eccentricity can be excited to large values ($>0.9$) when the initial inclination is less than $20^\circ$ for evection resonance, $ \sim 50^\circ$ for eviction resonance, $ \sim 140^\circ$ for $\sigma_1=2f_2+2\omega-2\Omega$ and $ \sim 120^\circ$ for $\sigma_1=2f_2+2\omega-2\Omega$ resonances. In addition to eccentricity, inclination can also vary significantly. For instance, binaries captured in Evection resonance experience eccentricity and inclination increase simultaneously at low initial inclinations. 

We then apply our results to binary blackholes in AGN disc. We focus on a binary composed of IMBH and a stellar mass blackhole. Using an ensemble of single-averaged simulations we list the following conclusions: 
\begin{itemize}
    \item Near circular binaries can be captured in precession induced resonances at a distance which increases with binary separation, mass of the companion and decreases with the mass of the central object and the quadrupole potential of the disk. It also depends on the inclination of the binary in a non-monotonic manner. The exact relationship is shown in Eqn. \ref{eqn:locreslese}. In our fiducial simulations, the resonances occur at a distance of $10^{2.5}-10^{5.5}$ AU from the supermassive blackhole.
    
    \item The maximum eccentricity achieved by a binary trapped in a precession induced resonance depends on the inclination of the binary at capture. This is due to two reasons. Firstly, the libration timescale depends on the inclination and it is possible for the libration timescale to be comparable to the gravitational merger timescale for certain inclination range. When this happens the eccentricity excitation is  suppressed. In addition, even without gravitational radiation, the maximum eccentricity a binary can attain is set by the constancy of adiabatic invariants for migrating binaries.
    
    \item Eccentricity excitation is possible only for binaries trapped in resonances with libration timescales shorter than the AGN disk lifetime, the gravitational merger timescale and the migration timescales in AGN disks. In our fiducial simulations, only closely separated binaries ($a_1 \sim 0.1$ AU) trapped in resonances corresponding to $\sigma_1=2f_2-2\omega-2\Omega$ and $\sigma_1=2f_2+2\omega-2\Omega$ experience significant eccentricity excitation ($e_{max}>0.5$). {The eccentricity excitation causes the binaries to merger faster thereby enhancing the merger rate of compact binaries in AGN disc.}

    \item Binaries sweeping through precession induced resonances would merge on migration timescale if the eccentricity of the binary is sufficiently excited. If the eccentricity excitation is not sufficient, the binary would be disrupted. In our fiducial simulations we find that binaries can merge 1-2 orders of magnitude faster if they migrate at the fastest rate which allows them to stay in resonance.
\end{itemize}
In addition, using an ensemble of N-body simulations, we find that the resonance capture probability is high in regions of parameter space where eccentricity can be excited to large values.  We also show that migrating binaries can be captured in multiple successive precession induced resonances. 

As a caveat, we use simple prescriptions for the quadrupole potential of circum-blackhole discs and migration of the binary orbits in AGN discs in this study. We also ignore binary hardening and eccentricity damping caused by the gas in the disk. Similar to our previous work \citep{bhaskar_black_2022}, we expect the eccentricity of the binary to be excited by resonance sweeping as long as the hardening and eccentricity damping timescales are longer than the libration timescale. Hydrodynamical simulations can be used to model the system in greater detail. For instance, using hydrodynamical simulations, \cite{yaping2021} have found that the eccentricities of blackhole binaries embedded in AGN discs can be excited in regime where evection resonances are important. It remains an open question to study the eviction-like resonances using hydrodynamical simulations in the future.
\begin{acknowledgments}
The authors thank the referee for comments that significantly improved the quality of the paper. GL and HB are grateful for the support by NASA 80NSSC20K0641 and 80NSSC20K0522. This work used the Hive cluster, which is supported by the National Science Foundation under grant number 1828187.  This research was supported in part through research cyberinfrastrucutre resources and services provided by the Partnership for an Advanced Computing Environment (PACE) at the Georgia Institute of Technology, Atlanta, Georgia, USA.
\end{acknowledgments}

\bibliography{ref.bib}{}
\bibliographystyle{aasjournal}
\end{document}